\documentclass[twocolumn, tighten, times, twocolappendix]{aastex63}

\graphicspath{{./}{figures/}}

\usepackage{amsmath}
\usepackage{booktabs}
\usepackage{longtable}
\begin{document}
	
	\title{RR Lyrae-based Distances for 39 Nearby Dwarf Galaxies Calibrated to Gaia eDR3}

	\author[0000-0002-1386-0603]{Pranav Nagarajan}
	\affiliation{Department of Astronomy, University of California, Berkeley, Berkeley, CA, 94720, USA}
	
	\author[0000-0002-6442-6030]{Daniel R. Weisz}
	\affiliation{Department of Astronomy, University of California, Berkeley, Berkeley, CA, 94720, USA}
	
	\author{Kareem El-Badry}
	\affiliation{Department of Astronomy, University of California, Berkeley, Berkeley, CA, 94720, USA}
	\affiliation{Center for Astrophysics $|$ Harvard \& Smithsonian, 60 Garden Street, Cambridge, MA 02138, USA}
	\affiliation{Harvard Society of Fellows, 78 Mount Auburn Street, Cambridge, MA 02138}

	
	
	\begin{abstract}
		We provide uniform RR Lyrae-based distances to 39 dwarf galaxies in and around the Local Group. We determine distances based on a Bayesian hierarchical model that uses periods and magnitudes of published RR Lyrae in dwarf galaxies and is anchored to well-studied Milky Way (MW) RR Lyrae with spectroscopic metallicities and Gaia eDR3 parallaxes. Gaia eDR3 parallaxes for the anchor sample are a factor of 2, on average, more precise than DR2 parallaxes, and allow for a much better constrained period-luminosity-metallicity relation. While $\sim75$\% of our distances are within 1-$\sigma$ of recent literature RR Lyrae distances, our distances are also $\sim2$-$3$ times more precise than literature distances, on average. On average, our distances are $\sim0.05$~mag closer than literature distances, as well as $\sim0.06$~mag closer than distances derived using a theoretical period-luminosity-metallicity relation. These discrepancies are largely due to our eDR3 parallax anchor. We show that the \textit{Hipparcos}-anchored RR Lyrae distance scale of \citet{carretta2000} over-predicts distances to MW RR Lyrae by $\sim0.05$~mag. The largest uncertainties in our distances are (i) the lack of direct metallicity measurements for RR Lyrae and (ii) the heterogeneity of published RR Lyrae photometry. We provide simple formulae to place new dwarf galaxies with RR Lyrae on a common distance scale with this work. We provide public code that can easily incorporate additional galaxies and data from future surveys, providing a versatile framework for cartography of the local Universe with RR Lyrae.

	\end{abstract}
	

	
	
	\section{Introduction} \label{sec:intro}
	
	Observational and theoretical studies of low-mass galaxies ($M_{\rm star} \lesssim 10^8\,M_{\odot}$) in and around the Local Group (LG) provide unique insight into a wide variety of astrophysics and cosmology including reionization, structure formation, star formation and chemical enrichment processes in low-mass halos, and the nature of dark matter \citep[e.g.,][]{mateo1998, tolstoy2009, kirby2011, mcconnachie2012,brown2014, weisz2014a, gallart2015, bullock2017, skillman2017, simon2019, mcconnachie2020, patel2020}.
	
	Central to virtually all of this science is accurate and precise knowledge of each galaxy's distance.   By virtue of their close proximities, galaxies in and around the LG can be resolved into individual stars, which allows for several ``gold standard'' distance indicators.  In a modest amount of integration time with ground- or space-based facilities, it is possible to resolve the tip-of-the-red giant branch (TRGB), horizontal branch (HB), main sequence turnoff (MSTO), and several classes of variable stars (e.g., RR Lyrae, Cepheid, Mira) in the same galaxy, providing several ways for determining a galaxy's distance (see \citealt{beaton2018} and references therein).  Indeed, a search of the literature (e.g., via the NASA Astrophysical Database; NED) reveals that many LG dwarf galaxies have dozens of distance determinations based on resolved stars, some of which trace back several decades \citep[e.g.,][]{shapley1938}.
	
\begin{figure}[t!]
		\centering
		\includegraphics[width = \columnwidth]{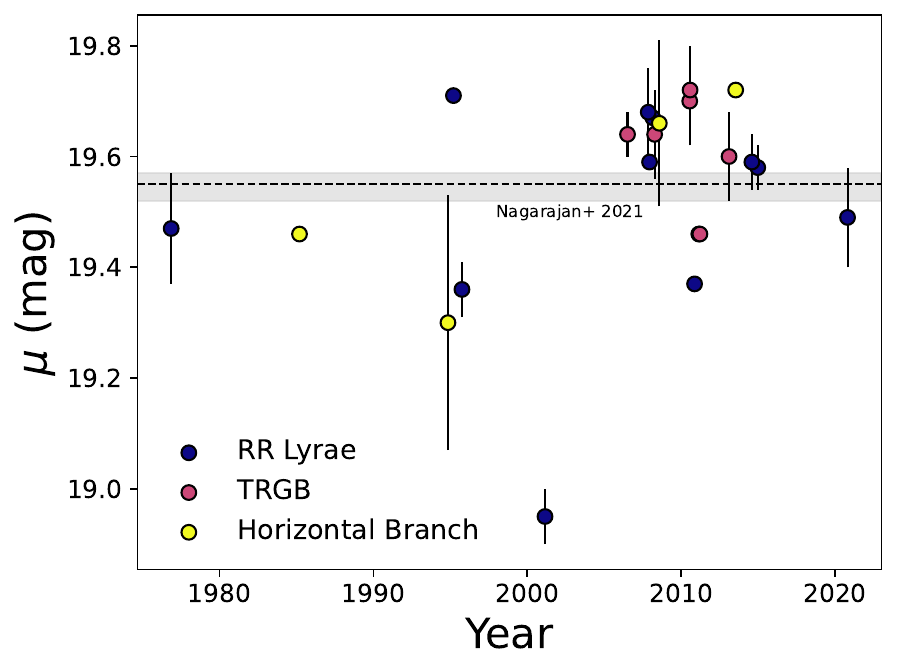}
		\caption{Select TRGB, HB, and RR Lyrae literature distances for Sculptor dSph. The convergence of various distances estimates to $\mu \sim19.60-19.70$ post-2000 is largely related to widespread use of \textit{Hipparcos} zero point calibrations \citep[e.g.,][]{carretta2000, rizzi2007}. The RR Lyrae distance measured in this paper ($\mu=19.55_{-0.03}^{+0.02}$, anchored to {\it Gaia} eDR3) is shown as the horizontal-dashed line, with 1-$\sigma$ uncertainty indicated by the grey band.}
		\label{fig:sculptor_distances}
	\end{figure}

	Despite (or perhaps, because of) the wealth of data and distance determination methods in the literature, the current state of distances to LG and nearby isolated low-mass galaxies is extremely heterogeneous.   As one example, NED reports 35 literature distances to the Sculptor Dwarf Spheroidal Galaxy, a list that is likely incomplete.  Even when limited to the modern era (i.e., CCD detectors after the year 2000), the distance to Sculptor dSph spans nearly a full magnitude in distance modulus\footnote{Here and in the rest of this work, we use ``distance modulus'' to refer to the quantity $\mu = 5\log\left( d/{\rm 10\,pc} \right)$, where $d$ is distance.} (e.g., \citealt{kovacs2001, huxor2015}), as illustrated in Figure \ref{fig:sculptor_distances}.  Narrowing this comparison to the same distance indicator (e.g., TRGB) and even more contemporary data (i.e., post-2010), the distance modulus to Sculptor varies from $\mu=19.46$ to $19.70$ (Figure \ref{fig:sculptor_distances}; e.g., \citealt{menzies2011, gorski2011, stringer_identifying_2021}). These published distances often use data of different quality, a variety of measurement techniques, and employ many underlying distance anchors. Such large variations and heterogeneity can severely affect science outcomes such as star formation histories (SFHs) and orbital histories. In an even broader context, dwarf galaxy distances are related to the ongoing ``Hubble tension'', which in part is predicated on high fidelity distances to select nearby galaxies \citep[e.g.,][]{beaton2018, riess2019, freedman2020, divalentino2021}.

	Beyond obfuscating the distance to an individual galaxy, the variable fidelity of distances in the literature can complicate comparative studies among multiple galaxies.  For example, more luminous ``classical'' dwarf galaxies often afford a precise TRGB distance determination by virtue of having a well-populated RGB.  In comparison, ultra-faint dwarfs (UFDs) are too sparsely populated for TRGB determinations, leaving RR Lyrae as the primary distance indicators.  While the two methods can be, in principle, consistently calibrated and used, much of the literature does not afford this degree of uniformity \citep[e.g.,][]{mcconnachie2012}, as many distance determinations result from studies of single or small sets of galaxies with different implicit zeropoints.  Alternatively, few existing sets of distances for larger samples of galaxies are limited to using sub-optimal distance indicators (e.g., the TRGB for ultra-faints, main sequence fitting) across the entire sample, resulting in poorly constrained distances for a fraction of the sample \citep[e.g.,][]{conn2012, brown2014}.  As a result, studies of multiple galaxies (e.g., for SFHs or orbital histories) that rely on distances, must select from a variety of heterogeneous and/or poorly constrained literature distances, which can result in hard-to-quantify systematics.

	Current empirical evidence suggests that RR Lyrae are present in all nearby dwarf galaxies for which accurate time series data are available \citep[e.g.,][]{martinez-vazquez_search_2019, stringer_identifying_2021}, implying they may be among the most promising route for homogeneous distances across the LG. RR Lyrae are warm ($T_{\rm eff}\sim 6,500\,\rm K$), low-mass $(M\lesssim 0.8 M_{\odot})$, relatively luminous ($L\sim 100\,L_{\odot}$) pulsating stars that have lost some of their hydrogen envelope on the first giant branch and are undergoing core helium burning \citep[e.g.][]{Iben1970}. RR Lyrae are found primarily in old, metal-poor stellar populations and are thus ideal tracers of globular clusters and dwarf galaxies \citep[e.g.][]{Preston1959, Sandage1970}. They follow a well-studied period-Wesenheit magnitude-metallicity relationship\footnote{See Section \ref{sec:wesenheit} for an explanation of Wesenheit magnitudes.} \citep[e.g.][]{vanAlbada1971, Breger1975, Longmore1986} that has recently been directly anchored to {\it Gaia} geometric parallaxes \citep[e.g.,][]{Neeley2019}, and a number of ground- and space-based observational campaigns have led to high-quality RR Lyrae data for dozens of nearby, low-mass galaxies.

	In this paper, we combine the wealth of literature RR Lyrae data for nearby dwarf galaxies, and updated parallaxes from \textit{Gaia} eDR3 for metal-poor RR Lyrae in the Milky Way (MW), with a hierarchical Bayesian approach in order to measure uniform RR Lyrae-based distances to 39 galaxies in and around the LG.  The final result is a set of distances that are on the same scale, anchored to geometric distances from \textit{Gaia}, and have uniformly determined uncertainties that includes co-variances between galaxy distances.  This approach and the resulting distances represent a step toward self-consistent distances for all low-mass galaxies in the local Universe that are anchored to a common and modern distance scale.

	This paper is organized as follows.  We describe the LG RR Lyrae data from the literature and the \textit{Gaia} RR Lyrae data and associated optical photometry in \S \ref{sec:data}.  In \S \ref{sec:method}, we describe the hierarchical model used to measure galaxy distances.  We apply our model and present the results in \S \ref{sec:results}.  We compare our results to published literature distances and distance anchors based on theoretical models of RR Lyrae, discuss caveats and assumptions, and provide updates to cartography of nearby galaxies in \S \ref{sec:discussion}.  In \S \ref{sec:conclusions} we summarize our main findings. We present tests on mock data and other technical verification details in the Appendices.

	\begin{figure*}[]
		\centering
		\includegraphics[width = \textwidth]{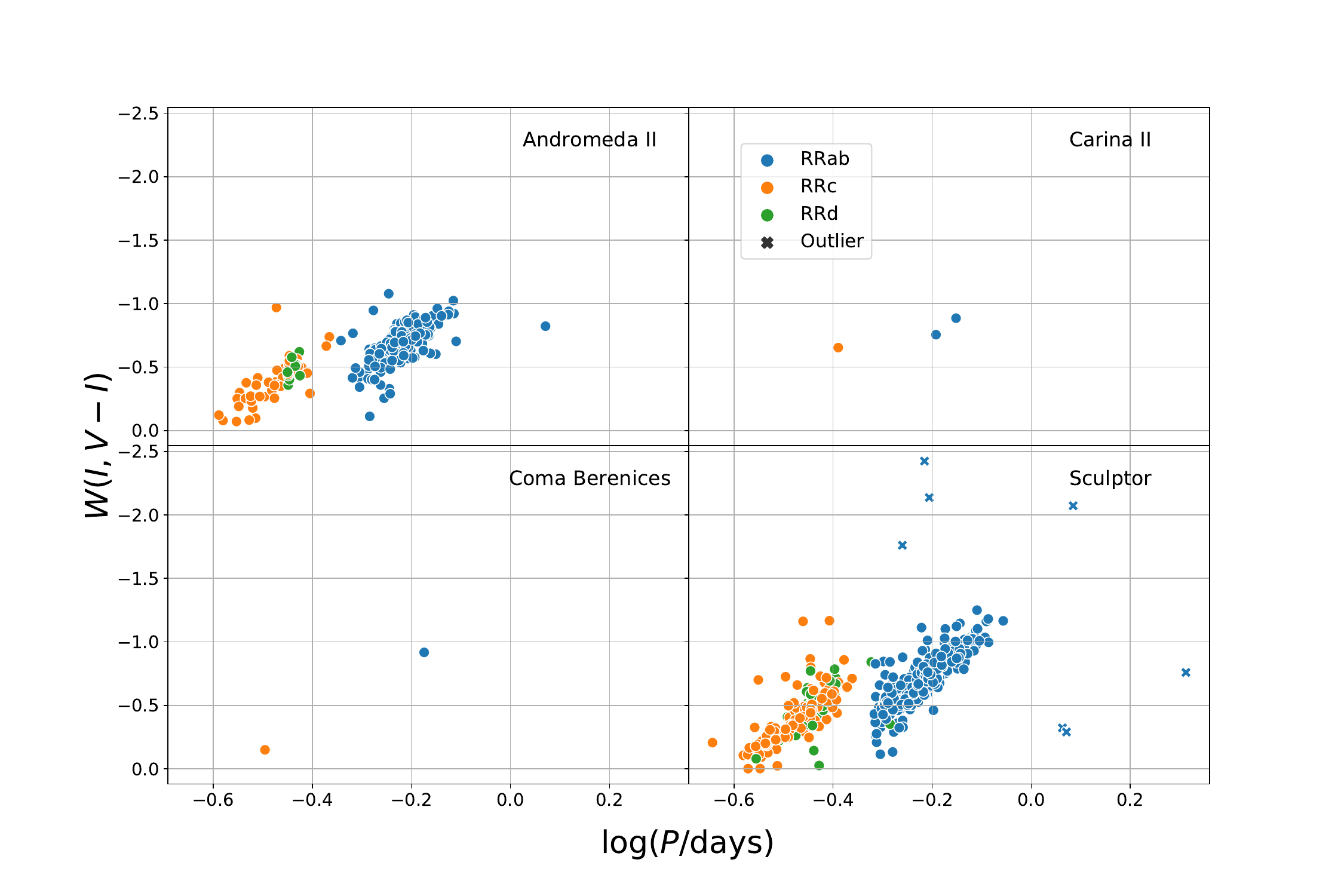}
		\caption{Examples of literature data for RR Lyrae from a few LG galaxies in our sample.  We plot the absolute Wesenheit magnitude (as computed in \S \ref{sec:data}) vs. the reported period for each RR Lyrae. Filled circles are color-coded by RR Lyrae type as reported in the literature.  Significant outliers (e.g., marginal RR Lyrae candidates) that were not used in literature analysis and are excluded from our analysis are marked with ``x'' symbols.  As discussed in \S \ref{sec:data}, we only model  RR Lyrae of type RRab in this paper. }
		\label{fig:examplerrlyrae}
	\end{figure*}
	
	\section{Data}\label{sec:data}
	
	\subsection{RR Lyrae in Local Group Dwarf Galaxies}
	
	\subsubsection{Sample Selection}
	
	This work leverages the rich set of optical RR Lyrae observations  throughout the LG that have been collected in the literature over the past decade, which we summarize in Table \ref{tab:lgdata}.  We draw our sample from the RR Lyrae table listed in Appendix A1 of \citet{martinez-vazquez_search_2019} with updates from a handful of recent papers \citep[e.g.,][]{stringer_identifying_2021}.
	
	For galaxies close to the MW, most observations of RR Lyrae are generally performed with ground-based facilities, either as single galaxy studies or larger photometric surveys (e.g., DES).  For galaxies beyond the MW halo ($d\gtrsim 100\,\rm kpc$), where RR Lyrae are faint and are often located in regions of modest-to-high levels of crowding, the Hubble Space Telescope (\textit{HST}) is the primary source of data.
	
	For inclusion in this study, we require that a previous study reported a period and mean magnitude for at least one RR Lyrae in a galaxy in at least two bands ($B$ and $V$, $V$ and $I$, or similar combinations that can be transformed), which allows us to account for extinction.  Thus, some galaxies with RR Lyrae-based distances reported in the literature are not included in our sample because not all necessary data could be collated (e.g., no reported periods, or only 1 band of published photometry).
	
	Due to the heterogeneity of the datasets (e.g., variations in light curve cadence, signal-to-noise, or foreground contamination), we first spot check the reported periods and mean magnitudes using our own code. For papers in which the light curves were published, we measured the periods for a subset ($\sim25$\%) of the RR Lyrae by determining the period of maximum power in (a) a Lomb-Scargle periodogram and (b) the ``hybrid'' periodogram from \citet{Saha2017}.  From this spot-checking exercise, we found that $\sim98$\% of the reported periods in the literature agreed with our recovered periods within $\sim1$\%.  The small number of disagreements could usually be attributed to cases of a few outliers in the light curves that were likely excluded from the literature analysis, but not clearly indicated as such in the papers.  By removing the outlying light curve points, we found periods that were consistent with those reported in the literature in $>99$\% of cases. The net result of this exercise is that we have high confidence in the periods reported in the literature for RR Lyrae in our sample.  
	
	We similarly spot-checked the mean magnitude calculations and found good consistency.  We thus make use of the  periods and mean magnitudes from the literature. We exclude several Andromeda satellites  with data in the B and V bands from our sample, due to the presence of a significant amount of scatter in the period-luminosity-metallicity relationship \citep[e.g., And~XIX, And~VII;][]{cusano2013, monelli_variable_2017}, which may be due to particularly noisy photometry or differential extinction in the case of And~VII.  The ongoing HST Treasury Program aimed at M31 satellites (HST-GO-15902; PI~D.~Weisz) will provide homogeneous RR Lyrae data for the entire M31 satellite population.
	
	Our final sample consists of a large, diverse population of dwarf galaxies that span the full range of dwarf galaxy morphology and factors of $\sim10^7$ in galaxy luminosity and $\sim60$ in mean stellar metallicity.  Though we originally limited our sample to galaxies with literature distances inside the LG ($\lesssim$1~Mpc), a handful of slightly more distant galaxies have published RR Lyrae data, motivating us to include them.  The most distant galaxy in our sample (ESO294-G010) has a literature distance of $\mu=26.40$ ($\sim1.9$~Mpc; \citealt{yang_early_2014}).    In total, out of the 39 galaxies in our sample, 8 galaxies are field dwarf galaxies, 10 are satellites of M31, and the remaining 21 galaxies orbit the MW.

	\subsubsection{Filters}
	\label{sec:filters}
	As shown in Table \ref{tab:lgdata}, the literature data we analyze includes (phase-averaged) apparent magnitudes in several filters.  However, optical data for the  \textit{Gaia} anchor sample, described below, is only uniformly available in Johnson-Cousins filters.  For consistency with the anchor sample, we transformed the mean magnitudes of the dwarf galaxy RR Lyrae from the observed bandpasses to either $V$ and $I$ or $B$ and $V$, whichever pair is closest in wavelength.  Such filter transformations are commonly employed for RR Lyrae studies and have been shown to be a very small source of uncertainty \citep[e.g.,][]{mcquinn_leo_2015, martinez-vazquez_islands_2017}.  We provide details on our adopted filter transformations in Appendix \S \ref{sec:filtertransforms}.
	
	\begin{deluxetable*}{ccccccc}
		\tablecaption{Summary of dwarf galaxy sample and their RR Lyrae. (1) Galaxy name; (2) Apparent $V$ magnitude of galaxy from the updated version of \citet{mcconnachie2012}; (3) Number of RRab stars used in model; (4) Total number of reported RR Lyrae in all classes; (5) Filters in which RR Lyrae were observed; (6) Transformed filters used in our model (see \S~\ref{sec:filters}); (7) Reference for RR Lyrae data.  If the observed and transformed filters are identical, no filter transformation was applied.\label{tab:lgdata}}
		\tablehead{\colhead{Galaxy} & \colhead{$m_V$} & \colhead{$N_{RRab}$} & \colhead{$N_{RRL}$} & \colhead{Observed} & \colhead{Transformed} & \colhead{References} \\
			\colhead{(1)} & \colhead{(2)} & \colhead{(3)} & \colhead{(4)} & \colhead{(5)} & \colhead{(6)}  & \colhead{(7)}}
		\startdata
		Andromeda I & 12.7 $\pm$ 0.1 & 229 & 296 & F475W, F814W & V, I & \citet{martinez-vazquez_islands_2017} \\
		Andromeda II & 11.7 $\pm$ 0.2 & 187 & 251 & F475W, F814W & V, I & \citet{martinez-vazquez_islands_2017} \\
		Andromeda III & 14.4 $\pm$ 0.3 & 84 & 111 & F475W, F814W & V, I & \citet{martinez-vazquez_islands_2017} \\
		Andromeda XI & 17.5 $\pm$ 1.2 & 10 & 15 & F606W, F814W & V, I & \citet{yang_hst/wfpc2_2012} \\
		Andromeda XIII & 18.1 $\pm$ 1.2 & 8 & 9 & F606W, F814W & V, I & \citet{yang_hst/wfpc2_2012} \\
		Andromeda XV & 14.6 $\pm$ 0.3 & 80 & 117 & F475W, F814W & V, I & \citet{martinez-vazquez_islands_2017} \\
		Andromeda XVI & 14.4 $\pm$ 0.3 & 3 & 8 & F475W, F814W & V, I & \citet{monelli_islands_2016} \\
		Andromeda XXVIII & 15.6 $\pm$ 0.9 & 35 & 85 & F475W, F814W & V, I & \citet{martinez-vazquez_islands_2017} \\
		Bootes I & 12.8 $\pm$ 0.2 & 5 & 15 & V, I & V, I & \citet{dallora_variable_2006} \\
		Canes Venatici I & 13.1 $\pm$ 0.2 & 18 & 23 & BVI & B, V & \citet{kuehn_variable_2008} \\
		Canes Venatici II & 16.1 $\pm$ 0.5 & 1 & 2 & BVI & B, V & \citet{greco_newly_2008} \\
		Carina I & 11.0 $\pm$ 0.5 & 62 & 83 & UBVI & V, I & \citet{coppola_carina_2015} \\
		Carina II & 13.3 $\pm$ 0.1 & 2 & 3 & gri & V, I & \citet{torrealba_discovery_2018} \\
		Cetus & 13.2 $\pm$ 0.2 & 147 & 172 & F475W, F814W & V, I & \citet{bernard_acs_2009} \\
		Coma Berenices & 14.1 $\pm$ 0.5 & 1 & 2 & BVI & V, I & \citet{musella_pulsating_2009} \\
		Crater II & 12.15 $\pm$ 0.02 & 28 & 34 & BVI & V, I & \citet{monelli_variable_2018}, \citet{vivas_walker_et_al_2019} \\
		Draco & 10.6 $\pm$ 0.2 & 211 & 267 & V, I & V, I & \citet{kinemuchi_variable_2008} \\
		Eridanus II & 15.74 $\pm$ 0.05 & 37 & 67 & gri & V, I & \citet{martinez-vazquez_monelli_cassisi_et_al_2021} \\
		ESO294-G010 & 15.3 $\pm$ 0.3 & 219 & 232 & F606W, F814W & V, I & \citet{yang_early_2014} \\
		ESO410-G005 & 14.9 $\pm$ 0.3 & 224 & 268 & F606W, F814W & V, I & \citet{yang_early_2014} \\
		Fornax & 7.4 $\pm$ 0.3 & 1386 & 1443 & ugriz & V, I & \citet{stringer_identifying_2021} \\
		Grus I & 17.1 $\pm$ 0.3 & 2 & 2 & gri & V, I & \citet{martinez-vazquez_search_2019} \\
		Hercules & 14.0 $\pm$ 0.3 & 6 & 12 & B, V & B, V & \citet{musella_stellar_2012} \\
		IC 1613 & 9.2 $\pm$ 0.1 & 61 & 90 & F475W, F814W & V, I & \citet{bernard_acs_2010} \\
		Leo A & 12.4 $\pm$ 0.2 & 7 & 10 & F475W, F814W & V, I & \citet{bernard_acs_2013} \\
		Leo I & 10.0 $\pm$ 0.3 & 68 & 164 & UBVRI & V, I & \citet{stetson_homogeneous_2014} \\
		Leo IV & 15.1 $\pm$ 0.4 & 3 & 3 & BVI & B, V & \citet{moretti_leo_2009} \\
		Leo P & 16.8 $\pm$ 0.2 & 9 & 10 & F475W, F814W & V, I & \citet{mcquinn_leo_2015} \\
		Leo T & 15.1 $\pm$ 0.5 & 1 & 1 & F606W, F814W & V, I & \citet{clementini_variability_2012} \\
		NGC 147 & 9.5 $\pm$ 0.1 & 113 & 177 & F606W, F814W & V, I & \citet{monelli_variable_2017} \\
		NGC 185 & 9.2 $\pm$ 0.1 & 531 & 820 & F606W, F814W & V, I & \citet{monelli_variable_2017} \\
		Phoenix I & 13.2 $\pm$ 0.4 & 95 & 121 & F555W, F814W & V, I & \citet{ordonez_rr_2014} \\
		Phoenix II & 17.3 $\pm$ 0.4 & 1 & 1 & gri & V, I & \citet{martinez-vazquez_search_2019}  \\
		Sagittarius II & 13.83 $\pm$ 0.1 & 1 & 5 & B, V & B, V & \citet{joo_rr_2019} \\
		Sculptor & 8.6 $\pm$ 0.5 & 271 & 536 & BVRI & V, I & \citet{martinez-vazquez_variable_2016} \\
		Segue II & 15.2 $\pm$ 0.3 & 1 & 1 & B, V & B, V & \citet{boettcher_search_2013} \\
		Tucana & 15.2 $\pm$ 0.2 & 216 & 358 & F475W, F814W & V, I & \citet{bernard_acs_2009} \\
		Ursa Major I & 14.4 $\pm$ 0.3 & 5 & 7 & B, V & B, V & \citet{garofalo_variable_2013} \\
		Ursa Major II & 13.3 $\pm$ 0.5 & 1 & 1 & BVI & B, V & \citet{dallora_stellar_2012} \\
		\enddata
	\end{deluxetable*}
	
	\subsubsection{Extinction and Wesenheit Magnitudes}
	\label{sec:wesenheit}
	
	We account for the effects of extinction and reddening by computing the Wesenheit magnitudes \citep[e.g.,][]{vandenBergh1968, Madore1976, madore1982} for RR Lyrae in each galaxy.  Wesenheit Magnitudes are ``reddening-free'' quantities in the sense that if two stars have the same intrinsic color and  magnitude, they will have the same Wesenheit magnitude, irrespective of the amount of extinction. They are constructed as:
	\begin{equation}
	W(M_1, M_2 - M_1) = M_1 - \alpha (M_2 - M_1),
	\end{equation}
	where $M_1$ and $M_2$ are magnitudes in two different bands; $M_1$ is a redder band than $M_2$. The color coefficient $\alpha$ depends on the two bands chosen:
	
	\begin{equation}
	\alpha = \frac{A_1}{A_2 - A_1},
	\end{equation}
	where the specific extinction $A_{\lambda}$ in a given band is computed as:
	
	\begin{equation}
	A_{\lambda} = R_{\lambda} E(B - V)
	\end{equation}
	and the extinction ratio $R_{\lambda}$ is adopted from known reddening laws.  In our case, we use the MW reddening law of \citet{cardelli1989} with $R_B = 4.151$, $R_V = 3.128$, and $R_I = 1.860$ as listed in Table 2 of \citet{Neeley2019}. Due to their low-metallicities and lack of recent star formation, most galaxies in our sample have little-to-no internal extinction \citep[e.g.,][]{dolphin2003}, particularly in regions in which the RR Lyrae are located.   Thus, we only account for foreground reddening from the MW. Correspondingly, we use $\alpha = 3.058$ to compute the $W(V,B-V)$ and $\alpha = 1.467$ for $W(I,V-I)$. We note that because of the color-dependence of $W$, it is important for our modeling that mean magnitudes be robustly measured (necessitating decent phase coverage in light curves); otherwise, color errors can translate to errors in Wesenheit magnitude.
	
	Figure \ref{fig:examplerrlyrae} shows the Period-Wesenheit magnitude (PW) relationship for four galaxies selected to illustrate the range of RR Lyrae properties (e.g., numbers, type of RR Lyrae) and data quality (e.g., scatter) of our sample.  The color-coding indicates the type of RR Lyrae as reported in the literature, while the ``x'' symbols indicate low confidence classifications and/or measurements as either indicated in the source papers or identified by visual inspection (i.e.\ we impose the equivalent of a loose, few $\sigma$, clipping on large outliers). We exclude these data points, along with RRab stars exhibiting the Blazhko effect (i.e.\ long-period modulation which leads to uncertainty in the period of pulsation), from our analysis. A full list of excluded outliers can be found on the project GitHub repository\footnote{\url{https://github.com/pranav-nagarajan/Mapping-Local-Group}}.

\begin{figure*}
		\centering
		\includegraphics[width=\textwidth]{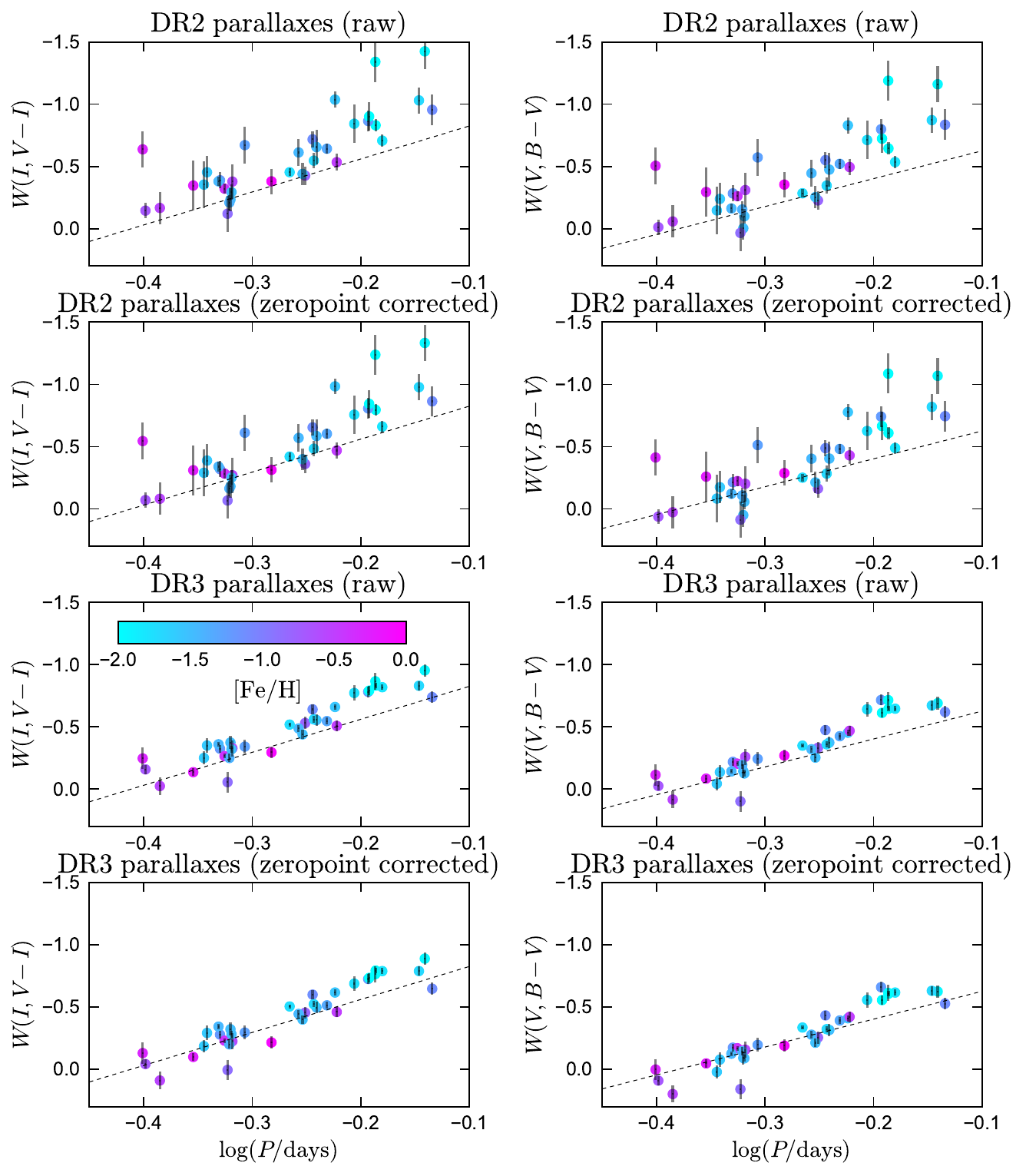}
		\caption{Milky Way RRab stars with high-precision {\it Gaia} parallaxes and spectroscopic metallicities. This sample anchors the PWZ we fit in dwarf galaxies to {\it Gaia} parallaxes. Left panels show $W(I, V-I)$ Wesenheit magnitudes; right panels show $W(V, B-V)$. In each case, we compare relations calculated with {\it Gaia} DR2 vs. eDR3 parallaxes, and those calculated with and without a parallax zeropoint correction. Dashed line is the same in all panels in a given column; it shows our best-fit relation for [Fe/H]=-1. Scatter in the PWZ due to parallax errors declined considerably from DR2 to eDR3. The metallicity trend is more obvious in the $W(I, V-I)$ relation.  }
		\label{fig:gaia}
	\end{figure*}

	For And~II and Sculptor, the two well-populated examples, we see clear trends between the periods and Wesenheit magnitudes, consistent with expectations (e.g., Figure \ref{fig:gaia}).  For Coma~Berenices (Coma~Ber) and Carina~II, the trends are not as clear due to the small number of RR Lyrae. Nevertheless, as we show, our model can provide robust distances to galaxies even with only handful of RR Lyrae.  We show the PW relationships for all galaxies in our sample  
	in Appendix~\ref{sec:all_galaxies}. 
	
	In order to calculate the errors on our Wesenheit magnitudes, we use standard error propagation. Assuming that magnitude $M_1$ has error $\sigma_1$ and magnitude $M_2$ has error $\sigma_2$, the error in $W(M_1, M_2 - M_1)$ is given by:
	
	\begin{equation}
	\sqrt{(1 + \alpha)^2 \sigma_1^2 + \alpha^2  \sigma_2^2}
	\end{equation}
	
	Due to the absence of reported uncertainties for most mean magnitudes in the literature, we must assume a typical uncertainty.  We adopted an uncertainty in the mean apparent magnitude of each RR Lyrae of 0.03 mag, which is typical of what is reported in the literature.  We discuss this assumption further in Section \ref{sec:method}, including the effect it has on the Wesenheit magnitudes.
	
	For simplicity, we measure distances only using RR Lyrae of type ``RRab'', i.e., fundamental pulsators.  While it is possible to ``fundamentalize'' the periods of RR Lyrae of other types (e.g., ``RRc''; first overtone pulsators) using simple transformations \citep[e.g.][]{vanAlbada1971,  vanAlbada1973}, the reliability of such transformations over a range of metallicities, ages, and mass loss rates is not well-established. In particular, it is not clear that the slopes of the period-luminosity relations for the two pulsation modes are the same. It is thus challenging to account for and propagate uncertainties introduced through fundamentalization.  We therefore select a purer sample, containing only RRab stars. Exploring using multiple types of RR Lyrae in tandem for distance determinations in our model is a topic for future study.

	\subsection{Milky Way calibration sample}
	
	\label{sec:gaia}
	In addition to RR Lyrae stars in dwarf galaxies, we simultaneously fit a sample of 36 well-studied Milky Way RRab stars within a few kpc of the Sun with parallaxes measured by {\it Gaia}. This sample was recently summarized by \citet[][their Table 6]{Neeley2017} and \citet[][their Table 1]{Neeley2019}. Stars in the sample have apparent magnitudes in the range $V=9.4-12.2$ (median $V=10.5$) and distances of 0.44 to 3.7 kpc (median $d \approx 0.9$\,kpc). Reddening values were taken from \citet{Feast2008} for most stars, or from \citet{Schlegel1998} for one star (BB Pup). These are generally modest (median $E(B-V)=0.04$). The metallicities were compiled by \citet{Fernley1998}, with a typical uncertainty of 0.15 dex. Recently, \citet{Gilligan2021} measured updated metallicities for a sample of Milky Way RR Lyrae that includes a majority of the stars in the \citet{Neeley2017} sample. We compared the metallicities reported by \citet{Fernley1998} and \citet{Gilligan2021} and found them broadly consistent (median absolute difference of 0.15 dex, comparable to the reported uncertainties for both samples). We thus use the \citet{Fernley1998} metallicities throughout. {\it Gaia} DR2 absolute magnitudes for the sample were explored by \citet{Neeley2019}. We cross-matched the sample with {\it Gaia} eDR3 \citep{Gaia2020}  using the \texttt{dr2\_neighbourhood} catalog in the {\it Gaia} archive. Wesenheit magnitudes for the anchor sample are provided by \citet{Neeley2019} and are computed as described in \S \ref{sec:data}.

	This sample is shown in Figure~\ref{fig:gaia}. In addition to comparing DR2 and eDR3 parallaxes (top vs. bottom), we show the effects of using the raw parallaxes (assuming $d=1/\varpi$; left) or first correcting for the parallax zeropoint (right). For DR2, we follow \citet{Neeley2019} in assuming a global zeropoint of 0.03 mas; for eDR3, we use the magnitude-, color-, and position-dependent zeropoint inferred by \citet{Lindegren2020zpt}. We also explored using distances incorporating a geometric prior rather than simple parallax inversion \citep[e.g.][]{Bailer-Jones2021}, but found the resulting distance differences negligible for this sample of bright and nearby sources.

	The improvement of {\it Gaia} parallax errors from DR2 to eDR3 is obvious. Although {\it Gaia} parallax uncertainties for most sources improved by only 30\% on average \citep{Lindegren2020}, the typical improvement for sources in this sample is more than a factor of 2, due to better control of systematics for bright sources. Even in eDR3, parallax uncertainties for stars of this brightness are typically underestimated by up to 30\% \citep{Elbadry2021, Zinn2021}. We inflate the reported eDR3 parallax uncertainties as recommended by \citet[][their Equation 16]{Elbadry2021}. For this sample, the typical uncertainty inflation is 15\%.

	The metallicity trend in the period-Wesenheit magnitude-metallicity (PWZ) is obvious in Figure~\ref{fig:gaia}, particularly with the eDR3 parallaxes.
	The effects of the parallax zeropoint are subtle but not negligible: for a distance of 1 kpc, the $\sim 0.02$ mas zeropoint translates to a 0.04 mag systematic difference in distance modulus, which is comparable to the final uncertainties in our inferred distances to the dwarf galaxies in our sample. We discuss the effects of the zeropoint further in Appendix~\ref{sec:zpt}.
	
	Finally, Figure \ref{fig:gaia} illustrates the range of metallicities covered by our sample.  This sample is ideal for providing tight constraints on an empirical PWZ as they span a broad range of metallicities.  This range overlaps with metal-poor RR Lyrae in the MW \citep[]{for_sneden_preston_2011, chadid_sneden_preston_2017, sneden_preston_chadid_adamow_2017, crestani_2021} as well as the few RR Lyrae in dwarf galaxies that have spectroscopic metallicity determinations \citep{clementini2005}.

	For ease of reference, we summarize the MW RRL in the calibration sample in Appendix \ref{sec:mwrrlyrae}, with updated parallaxes from {\it Gaia} eDR3, and corrected parallax errors based on \citet{Elbadry2021}.

	\section{A Hierarchical Model for Galaxy Distances}
	\label{sec:method}
	
	We employ a hierarchical Bayesian model to determine the distances to all galaxies simultaneously.  Our model is structured such that each galaxy $i$ has $j$ RR Lyrae.  The data for each star is a reported period and Wesenheit magnitude, both based on what is reported in the literature (e.g., Table \ref{tab:lgdata}).  
	
	We adopt a functional PWZ relationship with parameters that are anchored by \textit{Gaia} eDR3 calibration RR Lyrae.  As described below, we fit for these parameters by simultaneously including all RRab stars in all galaxies and the 36 RRab stars in the MW anchor sample. We include a scatter term that accounts for dispersion in the observed Wesenheit magnitudes at fixed period and metallicity that exceeds the adopted photometric errors; this could be the result of either underestimated uncertainties or intrinsic scatter in the PWZ (due, e.g., to the range of ages and/or masses of RR Lyrae in the sample).
	
	In total, we have the following free parameters in our model: the distance modulus to each galaxy $\mu_i$, the parameters ($a$, $b$, $c$) of the PWZ, as defined below, and the intrinsic scatter term, $\sigma$.  
	
	After adopting priors on each parameter as in Table \ref{tab:priors}, we use the No-U-Turn Hamiltonian Markov chain Monte Carlo sampling algorithm (NUTS; \citealt{Hoffman2011} ) implemented in \texttt{PyMC3} to sample the posterior distribution.  
	
	We now describe the specific model details, including choices in prior and implementation.

	\subsection{Model Assumptions}
	
	Our model adopts the following baseline assumptions:
	
	\begin{enumerate}
		\item All of the RR Lyrae (including the calibration stars in the Milky Way) follow the same PWZ, characterized by a period slope, metallicity slope, and zero-point.  Our model includes these as free parameters. 
		
		\item The RR Lyrae in each galaxy have metallicities that are drawn from the same metallicity distribution function (MDF).  As we discuss in \S \ref{sec:discussion}, the MDF for RR Lyrae is likely to differ significantly from the MDF of the entire galaxy, and from measured MDFs of individual red giants. In other words, we do not assume that the mean metallicity of the RR Lyrae follows the mass-metallicity relationship for dwarf galaxies established by red giant star spectroscopy \citep[e.g.,][]{kirby2013}.
		
		\item Each galaxy has a single well-defined distance modulus, $\mu_i$. Our model is not appropriate in large nearby galaxies like the LMC or Sagittarius, in which RR Lyrae can reveal detailed 3D structure \citep[e.g.,][]{haschke2012, klein2014}.
		
		\item The reported mean apparent magnitudes of all dwarf galaxy RR Lyrae have the same error of 0.03 mag.  A large plurality of literature studies do not report errors on mean magnitudes.  Based on studies that do report values, we adopt 0.03 mag as the uncertainty in each RR Lyrae's reported mean magnitude in each band, as this is a typical value listed.  Using the error propagation described in \S \ref{sec:wesenheit} and our adopted mean magnitude uncertainties, the typical uncertainty in Wesenheit magnitude is $\sim0.1$ mag.
		
		\item Errors on the periods are assumed to be negligibly small.  In cases when errors on periods in the literature are reported, they are $<$1\%; but a majority of studies do not report errors on the periods. We therefore do not include them in our model. 
	\end{enumerate}

	\subsection{Period-Wesenheit Magnitude-Metallicity Relationship}
	\label{sec:PWZ}

	Our model assumes the following commonly used PWZ for RR Lyrae:
	
	\begin{equation}
	W(M_1, M_2-M_1) = a + b \log P + c {\rm [Fe / H]}
	\label{eqn:pwz}
	\end{equation}
	
	\noindent where the period is measured in days and the metallicity, [Fe/H], is in dex.  $a$, $b$, and $c$ are free parameters that we allow to be constrained by the data. The metallicity of each RR Lyrae is also a free parameter.
	
	By adding distance modulus ($\mu = W(m_1, m_2-m_1) - W(M_1, M_2-M_1)$) to both sides, we can rewrite the PWZ in terms of apparent magnitude and the distance:
	
	\begin{equation}
	W(m_1, m_2-m_1) = \mu + a + b \log P + c {\rm [Fe / H]}
	\end{equation}
	
	\noindent where $W(m_1, m_2-m_1)$ is the apparent Wesenheit magnitude for an RR Lyrae star.
	
	\subsection{Simultaneously Fitting RR Lyrae Distances and Metallicities in Multiple Galaxies}
	
	Our model supposes that we have $i$ galaxies and that each galaxy has $j$ RR Lyrae.  This hierarchical model is expressed as:
	
	\begin{equation}
	W(m_1, m_2-m_1)_{i, j} = \mu_{i} + a + b \log P_{i, j} + c {\rm [Fe / H]}_{i, j}
	\end{equation}
	where $\mu_i$ is the distance modulus to the $i$th galaxy, and $W(m_1, m_2-m_1)_{i, j}$, $\log P_{i, j}$, and ${\rm  [Fe / H]}_{i, j}$ are the apparent Wesenheit magnitude, log period, and metallicity of the $j$th RR Lyrae in the $i$th galaxy. 
	
	The values of $\log P_{i, j}$ and $W(m_1, m_2-m_1)_{i, j}$ for each RR Lyrae are based on what is reported in the literature (Table \ref{tab:lgdata}) with values for $W(m_1, m_2-m_1)_{i, j}$ calculated as described in \S \ref{sec:data}. The values of $\mu_i$ and $[Fe / H]_{i, j}$ (and their associated uncertainties) are well-constrained for each calibration star as described in \S \ref{sec:gaia} and Appendix \ref{sec:mwrrlyrae}.
	
	The likelihood function for an observed Wesenheit magnitude of a RR Lyrae variable is given by:
	
	\begin{equation}
	W(m_1, m_2 - m_1)_{i, j} \sim \mathcal{N}(\hat{W}(m_1, m_2 - m_1)_{i, j}, \sigma')
	\end{equation}
	
	where $\hat{W}(m_1, m_2 - m_1)_{i, j}$ is the predicted Wesenheit magnitude based on our model and $\sigma'$ is the quadrature sum of the intrinsic scatter $\sigma$ and the observational error $\sigma_{obs}$.

	The main unknowns for each galaxy are the underlying metallicities of the RR Lyrae and the distance moduli to individual galaxies. We adopt a (very) weakly informative prior on the distance modulus for each galaxy:
	
	\begin{equation}
	\mu_i \sim \mathcal{N} (20, 10),
	\end{equation}
	where $\mathcal{N}(x,\sigma)$ denotes a normal distribution with mean $x$ and variance $\sigma^2$.

	\begin{deluxetable}{ccc}
		\tablecaption{Prior distributions used in the hierarchical Bayesian model. Note that Normal distributions are expressed as Normal($\mu, \sigma$).  Note that the B-V and V-I samples each have their own PWZ and intrinsic scatter variables, though all have the same priors.  }
		\label{tab:priors}
		\tablehead{\colhead{Parameter} &
			\colhead{Prior} & \colhead{Description}}
		\startdata
		$a$ & $\mathcal{N}(0, 1)$ & Zeropoint \\
		$b$ & $\mathcal{N}(0, 1)$ & Period slope \\
		$c$ & $\mathcal{N}(0, 1)$ & Metallicity slope \\
		$\mu_i$ & $\mathcal{N}(20, 10)$ & Distance modulus of galaxy $i$ \\
		${\rm [Fe/H]}_{i, j}$ & $\mathcal{N}$($\langle \rm  [Fe/H] \rangle$, 0.5) & Metallicity of RR Lyrae $j$ in galaxy $i$ \\
		$\sigma$ & Half-Normal(0.5) & Intrinsic scatter \\
		\enddata
	\end{deluxetable}
	
	A number of more luminous dwarf galaxies have reported MDFs in the literature \citep[e.g.,][]{kirby2011}.  However, (a) not all dwarf galaxies, particularly those with very few bright red giants, have measured MDFs \citep[e.g.,][]{simon2019}; and (b) the reported MDFs come from red giants and are subjected to a variety of selection effects, meaning they may not be representative of the true MDFs of RR Lyrae. In most cases, we expect the red giant-based MDFs to be biased to higher metallicities compared to the RR Lyrae \citep[e.g.,][]{clementini2005, kirby2009}, which preferentially sample the oldest and most metal-poor stellar population.
	
	Because there are few dwarf galaxies in which RR Lyrae metallicities have been measured directly \citep[e.g.,][]{clementini2005}, we adopt the following approximation for the assumed MDFs in this work.  We assume that the metallicities for the RR Lyrae in each galaxy are drawn from a Gaussian MDF, where the mean RR Lyrae metallicity depends on the  value of $\langle \rm [Fe/H] \rangle$ reported in the literature (e.g., \citealt{mcconnachie2012} and subsequent updates). If $\langle \rm [Fe/H] \rangle < -2.00$, we adopt that value as the mean metallicity.   Otherwise, if the mean metallicity is more metal-rich than $-2.00$, we set  $\langle \rm [Fe/H] \rangle_{RRL} = -2.00$.  While it is unlikely that RR Lyrae in all LG dwarf galaxies have the same mean metallicity, we find this to be a more realistic assumption than e.g., adopting the mean metallicity reported in the literature or predicted by a galaxy mass-metallicity relation, since the specific frequency of RR Lyrae is expected to be much higher in the galaxies' oldest and most metal-poor populations \citep[e.g.][]{Suntzeff1991, Baker2015}. We discuss the rationale for and limitations of this assumption further in \S \ref{sec:discussion}.

	In addition to the mean metallicity, we also adopt the same MDF width of 0.5~dex for each galaxy, which is typical of measured dwarf galaxy MDFs in the literature \citep[e.g.,][]{kirby2013, simon2019}.  In Appendix~\ref{sec:mockdata}, we used mock data to test the sensitivity of our inferred distances to assumed MDF width. We find that modest changes ($\pm0.2$~dex) in the adopted MDF width introduce percent-level changes in the inferred distance moduli, verifying that this assumption is a subdominant source of uncertainty.

	\section{Galaxy Distances}
	\label{sec:results}
	
	\begin{deluxetable*}{cccccc}
		\tablecaption{Inferred PWZ Parameters. We report the median values of the posterior distributions, with the lower and upper error bounds computed using differences from the 16th and 84th percentiles respectively. Note that the linear regression models for the calibration star sample were fit around the mean log period of $-0.30$ and mean metallicity of $-1.36$ (Equation~\ref{eq:pwz}) in accordance with \citet{Neeley2019}. The zero point values have been adjusted to match the \citet{Neeley2019} fitting scheme (see \S \ref{sec:anchor}) for ease of comparison with the parameters from the full PWZ relations.}
		\label{tab:mwrrl}
		\tablehead{\colhead{Parameter} &
			\colhead{Description} & \colhead{$V$-$I$ galaxies} & \colhead{$B$-$V$ galaxies} & \colhead{MW anchor sample ($V$-$I$)} & \colhead{MW anchor sample ($B$-$V$)} \\
			\colhead{(1)} & \colhead{(2)} & \colhead{(3)} & \colhead{(4)} & \colhead{(5)} & \colhead{(6)}}
		\startdata
		$\sigma$ & Intrinsic scatter & $0.10^{+0.01}_{-0.01}$ & $0.05^{+0.02}_{-0.02}$ & $0.04^{+0.01}_{-0.01}$ & $0.04^{+0.02}_{-0.02}$ \\
		$a$ & Zeropoint & $-0.94^{+0.04}_{-0.05}$ & $-1.00^{+0.08}_{-0.08}$ & $-1.05^{+0.06}_{-0.06}$ & $-1.06^{+0.08}_{-0.07}$ \\
		$b$ & Coefficient of $\log{P}$ & $-2.66^{+0.05}_{-0.05}$ & $-2.79^{+0.21}_{-0.20}$ & $-2.77^{+0.19}_{-0.19}$ & $-2.96^{+0.21}_{-0.21}$ \\
		$c$ & Coefficient of $\rm [Fe/H]$ & $0.17^{+0.03}_{-0.03}$ & $0.03^{+0.03}_{-0.02}$ & $0.11^{+0.02}_{-0.02}$ & $0.02^{+0.03}_{-0.02}$ \\
		\enddata
	\end{deluxetable*}
	
	\subsection{Evaluating the model}
	
	Having defined our model, including priors on all parameters (Table \ref{tab:priors}), we sample from the model using \texttt{PyMC3's} NUTS sampler with 1000 tuning steps and 1000 draws.  We use the Gelman-Rubin convergence diagnostic \citep[$\hat{R}$;][]{Gelman1992} to assess convergence.  We also run the model on mock data and on the Milky Way calibration sample (without including any dwarf galaxy RR Lyrae) to ensure that the model behaves as expected and yields unbiased distances. These tests are detailed in Appendix~\ref{sec:mockdata}.
	
	Because the LG data do not all use the same filter combinations, we split the sample into two subsets, $B$-$V$ and $V$-$I$ galaxies, depending on which photometry is available (see Table~\ref{tab:lgdata}).  We use the same anchor sample of MW RR Lyrae (for which $BVI$ photometry is uniformly available) for both sets, but each set is modeled independently and has its own PWZ relation.
	
	\subsection{Results from fitting the anchor sample alone}
	\label{sec:anchor}
	
	We first fit only the $B$-$V$ and $V$-$I$ anchor samples without including any dwarf galaxy RR Lyrae, in order to compare our results with previous fits of the same with {\it Gaia} DR2 parallaxes \citep{Neeley2019}.  Specifically, we fit the 36 RRab stars with the model:

	\begin{equation}
	\begin{split}
	\label{eq:pwz}
	W(m_1, m_2 - m_1)_{i} = \mu_{i} + a' + b ~ (\log P_{i} + 0.30) & \\ + c ~ ( {\rm  [Fe / H]}_{i} + 1.36),
	\end{split}
	\end{equation}
	where $W(m_1, m_2 - m_1)_i$, $\mu_i$, $\log P_i$, and $[Fe / H]_i$ are the observed Wesenheit magnitude, distance modulus, log period, and metallicity for the $i$th calibration star. We follow \citet{Neeley2019} in using the mean values of log period ($-0.30$) and metallicity ([Fe/H] $= -1.36$) as ``pivot points'' in fitting, as this reduces the covariance between parameters of the PWZ and allows for more straightforward comparison to the \citet{Neeley2019} constraints.  More explicitly, we have adjusted the zero point values such that $a = a' + 0.30 b + 1.36 c$. Note that when fitting the entire sample (i.e., dwarf galaxies and anchor stars) we do not fit around a mean value; this is only done for consistency checks with \citet{Neeley2019}.
	
	Table \ref{tab:mwrrl} lists the results of our fits, and the dashed lines in Figure \ref{fig:gaia} shows our derived PW in the case of $[Fe/H] = -1$.
	
	Our PWZ parameters are all consistent with those reported in \citet{Neeley2019}. Each parameter is constrained to better than 10\%. Because the {\it Gaia} eDR3 parallaxes are more precise than those in DR2, our PWZ is also more tightly constrained than that found by \citet{Neeley2019}. Quantitatively, while the scatter around the fitted $I, V-I$ and $V, B-V$ PWZ relations found by \citet{Neeley2019} is 0.18, the intrinsic scatter around our fitted relations is 0.04, which is 4.5 times smaller. We note that the $\sigma$ provided by \citet{Neeley2019} is an RMS value calculated based on their best fit model, which is technically a different treatment of uncertainty in fit compared to our intrinsic scatter term. 
	
	For direct comparison with \citet{Neeley2019}, we also computed the RMS around the linear fit specified by our derived parameters. We found values of $0.076$ and $0.082$ for the $I, V-I$ and $V, B-V$ PWZ relations, respectively. These RMS values are more than 2 times smaller than those provided by \citet{Neeley2019}.
	
	Following a suggestion from the referee, we changed our assumed solar metallicity ($Z/X = 0.0274$) to match the value provided by \citet{asplund_2009} ($Z/X = 0.0181$), and used the delta S offset from \citet{crestani_2021} to place the RR Lyrae calibration stars on their newer metallicity scale. We re-ran our model and found that the derived parameters did not significantly change (i.e.,\ the two sets of derived parameters were consistent to within 1-$\sigma$), so we have not adopted these changes in our analysis.

	\subsection{Results from simultaneously fitting the anchor sample and LG dwarf galaxies}
	\label{sec:fitting_results}
	
	As described in \S \ref{sec:PWZ}, we simultaneously fit the anchor sample of 36 MW RR Lyrae and all RRab stars in each dwarf galaxy, treating the $B$-$V$ and $V$-$I$ samples separately. The free parameters of the fit are the parameters of the PWZ relation, the intrinsic scatter, the distance modulus of each dwarf galaxy, and the metallicity of each dwarf galaxy RR Lyrae.

\begin{figure*}[]
		\centering
		\includegraphics[width = \textwidth]{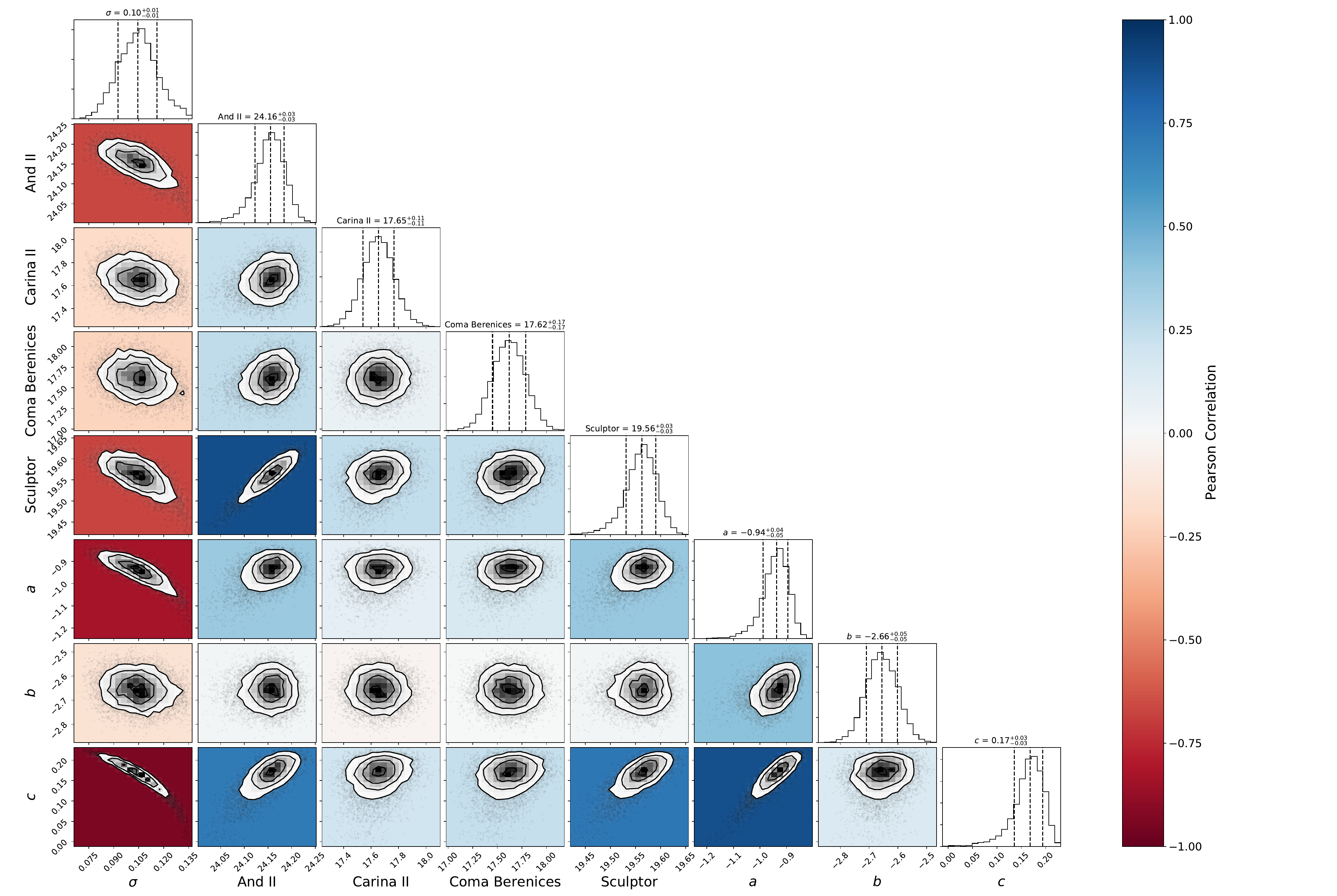}
		\caption{Corner plot from the final run of our MCMC sampling, only showing selected galaxies for clarity. The diagonal entries display the marginal distribution of each derived parameter, with dashed lines indicating the median value and 68th percentile constraints. Each of the other panels displays a joint distribution, with contours identifying regions of high sampling density in parameter space. Panels are color-coded by the covariance between pairs of parameters (see \S~\ref{sec:fitting_results} for discussion). Similar constraints for the full sample of galaxies can be found in Figure~\ref{fig:cornerVI}. }
		\label{fig:smallVI}
	\end{figure*}

	Figure \ref{fig:smallVI} shows the posterior distributions for the 4 example galaxies highlighted in Figure \ref{fig:examplerrlyrae}. Note that we fit all galaxies in the $V$-$I$ and $B$-$V$ subsets simultaneously, so the posteriors shown are only for a subset of the free parameters being fit. We have marginalized over the metallicities of the individual RR Lyrae in each galaxy, and the distance moduli of the other galaxies.
	
	The posterior constraints are generally unimodal and otherwise well-behaved. For each joint distribution, we compute the Pearson correlation coefficient and indicate it by the coloring of the plot's background. Dark blue indicates a significant positive correlation between two parameters and dark red indicates a significant negative correlation. The intrinsic scatter term, $\sigma$, is inversely correlated with the metallicity slope, $c$. This can be understood as follows. The metallicities of individual RR Lyrae in dwarf galaxies are not measured. When there is a range of apparent Wesenheit magnitudes in a galaxy at fixed period, this could either be due to a range of metallicities (if the metallicity slope, $c$, is large) or to a significant intrinsic scatter (if $c$ is small). This would lead to near-complete degeneracy between $\sigma$ and $c$ if we did not also include the MW anchor sample, for which spectroscopic metallicities are known, in the fit. 
	
	Because $c$ is positive (metal-poor RR Lyrae are brighter at fixed period), it is positively correlated with distance modulus for all galaxies. That is, if the dwarf galaxy RR Lyrae were more metal-poor, or the metallicity slope were steeper, the galaxies would be at larger inferred distance moduli. As a result of the anti-correlation between $c$ and $\sigma$, and the positive correlation between $c$ and $\mu_i$, $\sigma$ is anti-correlated with all distance moduli $\mu_i$. Because all galaxies are linked through the PWZ, the inferred distance moduli of individual galaxies are also often covariant.
	
	Table \ref{tab:mwrrl} lists our derived PWZ parameters and Table \ref{tab:final} contains the corresponding distance moduli for all galaxies. In all cases, our point estimates represent the median of the marginalized posterior distributions and the reported uncertainties reflect the 16th and 84th percentiles. 
	
	In general, we find the precision of our constraints to be consistent with expectations set by modeling simulated data. For example, compared to the mock data tests presented in Appendix \ref{sec:mockdata}, we find that the uncertainties in our reported values of the PWZ, scatter, and distances are comparable to what the mock data predict, with tighter constraints on the distances to galaxies with larger numbers of RR Lyrae. Similarly, we find that the derived PWZ parameters for the 30 $V$-$I$ galaxies plus the anchor sample are consistent with the derived PWZ parameters for the anchor-only sample, though the anchor-only sample is constrained less precisely, owing to a smaller number of RR Lyrae. The main exception is the intrinsic scatter, $\sigma$, which is significantly larger for the full $V$-$I$ sample than for the anchor sample alone. This suggests that underestimated photometric uncertainties (due, e.g., to crowding or poor light curve phase coverage) contributes non-negligibly to the scatter in the dwarf galaxy data. 
	
	The observed PW relationships for the full $V$-$I$ sample show significant variation in apparent data quality (Figure \ref{fig:cornerVI}). Some galaxies have tight linear PWs (e.g., IC~1613, Carina, Cetus), whereas others (e.g., Leo~I, Fornax, ESO294-G010) have modest to large scatter in $W(I,V-I)$ at fixed period. The scatter term is designed to capture this variability, which may explain why it is higher when the galaxies are fit alongside the anchor sample.  Other effects due to the heterogeneous nature of the data (e.g., filter transformations, inconsistent flux zeropoints among the studies) may also contribute.  The scatter in the $B$-$V$ sample is likely lower as the anchor sample has a much larger statistical weight on the overall fit.

	\subsection{Uniform distances to LG dwarf galaxies}

	Table \ref{tab:final} lists our {\it Gaia} eDR3-anchored distances to 39 LG galaxies.  Here, we report the median distance for each galaxy as determined from the marginalized posterior distributions for each galaxy.  The reported uncertainties reflect the 68\% confidence interval determined from the marginalized distributions. In general, we find that the distances to most galaxies are well-constrained and follow expectations set by our mock data tests presented in Appendix \ref{sec:mockdata}.
	
	The median distance modulus uncertainty across our sample is $\pm$0.07 mag, corresponding to a fractional distance uncertainty of about 3\%. The smallest uncertainties ($\pm$0.03~mag) correspond to the galaxies with the largest number of RR Lyrae (e.g., Draco, Sculptor), and the largest,  ($\pm$0.13~mag) to galaxies with a handful, or only one, RR Lyrae.

	\begin{longrotatetable}
		\begin{deluxetable*}{cccccccccc}
			\tablecaption{Summary of measured distances. (1) Galaxy; (2) {\it Gaia}-anchored distance modulus; (3) Literature distance modulus; (4) Theory-anchored distance modulus, (5) {\it Gaia}-anchored distance from Sun; (6) Literature distance from Sun; (7) 3D distance from MW center \citep{blandhawthorn2016}; (8) 3D distance from M31 assuming $\mu_{\rm M31}=24.47$ \citep{mcconnachie2012}; (9) Residual significance of {\it  Gaia}-anchored and literature distances in units of $\sigma$, (10) Assumed mean metallicity of RR Lyrae. $^{a}$ Despite having \textit{HST}-based RR Lyrae light curves published by \citet{monelli_variable_2017}, that paper did not derive distances from RR~Lyrae and instead quoted distances from various literature sources that used other methods.  We therefore list their quoted distances for completeness, but do not read too much into the level of agreement.\label{tab:final}}
			\tablehead{\colhead{Name} & \colhead{$\mu_{\text{This Work}}$} & \colhead{$\mu_{\text{Literature}}$} & \colhead{$\mu_{\text{Theory}}$} & \colhead{$d_{\text{This Work}}$} & \colhead{$d_{\text{Literature}}$} & \colhead{$d_{\text{MW}}$} &
				\colhead{$d_{\text{M31}}$} & \colhead{Literature difference} & \colhead{$\langle \rm [Fe/H] \rangle_{\text{Assumed}}$} \\
				\colhead{} & \colhead{(mag)} & \colhead{(mag)} & \colhead{(mag)} & \colhead{(kpc)} & \colhead{(kpc)} & \colhead{(kpc)}  & \colhead{(kpc)} & \colhead{($\sigma$)} & \colhead{(dex)} \\
				\colhead{(1)} & \colhead{(2)} & \colhead{(3)} & \colhead{(4)} & \colhead{(5)} & \colhead{(6)}  & \colhead{(7)} & \colhead{(8)} & \colhead{(9)} & \colhead{(10)}}
			\startdata
			Andromeda I & $24.47^{+0.02}_{-0.03}$ & $24.49^{+0.08}_{-0.08}$ & $24.53^{+0.01}_{-0.01}$ & $785^{+10}_{-12}$ & $791^{+29}_{-29}$ & $787^{+7}_{-11}$ & $45^{+1}_{-2}$ & 0.186 & -2.00 \\
			Andromeda II & $24.15^{+0.02}_{-0.03}$ & $24.16^{+0.08}_{-0.08}$ & $24.21^{+0.01}_{-0.01}$ & $678^{+9}_{-10}$ & $679^{+25}_{-25}$ & $681^{+6}_{-9}$ & $164^{+3}_{-5}$ & 0.052 & -2.00 \\
			Andromeda III & $24.34^{+0.03}_{-0.03}$ & $24.36^{+0.08}_{-0.08}$ & $24.40^{+0.01}_{-0.01}$ & $739^{+11}_{-12}$ & $745^{+27}_{-27}$ & $741^{+10}_{-10}$ & $76^{+5}_{-5}$ & 0.194 & -2.00 \\
			Andromeda XI & $24.38^{+0.05}_{-0.05}$ & $24.33^{+0.05}_{-0.05}$ & $24.45^{+0.04}_{-0.04}$ & $753^{+20}_{-20}$ & $735^{+17}_{-17}$ & $755^{+17}_{-17}$ & $103^{+3}_{-3}$ & 0.691 & -2.00 \\
			Andromeda XIII & $24.44^{+0.06}_{-0.06}$ & $24.62^{+0.05}_{-0.05}$ & $24.52^{+0.05}_{-0.05}$ & $776^{+22}_{-23}$ & $839^{+19}_{-19}$ & $776^{+21}_{-21}$ & $114^{+1}_{-1}$ & 2.151 & -2.00 \\
			Andromeda XV & $24.39^{+0.03}_{-0.03}$ & $24.42^{+0.08}_{-0.08}$ & $24.45^{+0.01}_{-0.01}$ & $756^{+11}_{-12}$ & $766^{+28}_{-28}$ & $760^{+10}_{-10}$ & $94^{+2}_{-2}$ & 0.310 & -2.00 \\
			Andromeda XVI & $23.66^{+0.09}_{-0.1}$ & $23.72^{+0.09}_{-0.09}$ & $23.72^{+0.09}_{-0.08}$ & $540^{+23}_{-25}$ & $555^{+23}_{-23}$ & $544^{+22}_{-25}$ & $260^{+19}_{-22}$ & 0.454 & -2.1 \\
			Andromeda XXVIII & $24.41^{+0.03}_{-0.04}$ & $24.43^{+0.08}_{-0.08}$ & $24.48^{+0.02}_{-0.02}$ & $764^{+13}_{-14}$ & $769^{+28}_{-28}$ & $762^{+11}_{-14}$ & $371^{+2}_{-3}$ & 0.174 & -2.00 \\
			Bootes I & $19.18^{+0.08}_{-0.08}$ & $19.11^{+0.08}_{-0.08}$ & $19.22^{+0.07}_{-0.06}$ & $69^{+3}_{-3}$ & $66^{+2}_{-2}$ & $66^{+3}_{-3}$ & $814^{+1}_{-1}$ & 0.636 & -2.55 \\
			Canes Venatici I & $21.49^{+0.04}_{-0.04}$ & $21.62^{+0.06}_{-0.06}$ & $21.51^{+0.04}_{-0.04}$ & $199^{+4}_{-4}$ & $211^{+6}_{-6}$ & $198^{+4}_{-4}$ & $848^{+2}_{-2}$ & 1.737 & -2.00 \\
			Canes Venatici II & $21.23^{+0.16}_{-0.16}$ & $21.02^{+0.06}_{-0.06}$ & $21.23^{+0.17}_{-0.17}$ & $176^{+13}_{-13}$ & $160^{+4}_{-4}$ & $177^{+13}_{-13}$ & $837^{+6}_{-6}$ & 1.224 & -2.21 \\
			Carina I & $19.98^{+0.03}_{-0.03}$ & $20.07^{+0.08}_{-0.08}$ & $20.05^{+0.01}_{-0.01}$ & $99^{+2}_{-2}$ & $103^{+4}_{-4}$ & $101^{+1}_{-1}$ & $831^{+1}_{-1}$ & 0.947 & -2.00 \\
			Carina II & $17.65^{+0.11}_{-0.11}$ & $17.86^{+0.02}_{-0.02}$ & $17.72^{+0.10}_{-0.11}$ & $34^{+2}_{-2}$ & $37^{+0}_{-0}$ & $35^{+2}_{-2}$ & $799^{+1}_{-1}$ & 1.800 & -2.00 \\
			Cetus & $24.43^{+0.02}_{-0.03}$ & $24.46^{+0.12}_{-0.12}$ & $24.49^{+0.01}_{-0.01}$ & $770^{+10}_{-12}$ & $780^{+43}_{-43}$ & $770^{+7}_{-11}$ & $683^{+3}_{-5}$ & 0.230 & -2.00 \\
			Coma Berenices & $17.62^{+0.16}_{-0.16}$ & $18.13^{+0.08}_{-0.08}$ & $17.68^{+0.15}_{-0.15}$ & $33^{+3}_{-3}$ & $42^{+2}_{-2}$ & $35^{+2}_{-2}$ & $791^{+1}_{-1}$ & 2.754 & -2.6 \\
			Crater II & $20.21^{+0.04}_{-0.04}$ & $20.3^{+0.08}_{-0.08}$ & $20.28^{+0.03}_{-0.02}$ & $110^{+2}_{-2}$ & $115^{+4}_{-4}$ & $109^{+2}_{-2}$ & $877^{+2}_{-2}$ & 0.940 & -2.00 \\
			Draco & $19.39^{+0.02}_{-0.03}$ & $19.58^{+0.15}_{-0.15}$ & $19.46^{+0.01}_{-0.01}$ & $76^{+1}_{-1}$ & $82^{+6}_{-6}$ & $76^{+1}_{-1}$ & $748^{+0}_{-0}$ & 1.200 & -2.00 \\
			Eridanus II & $22.67^{+0.04}_{-0.04}$ & $22.84^{+0.05}_{-0.05}$ & $22.73^{+0.02}_{-0.02}$ & $343^{+7}_{-8}$ & $370^{+9}_{-9}$ & $344^{+6}_{-6}$ & $870^{+3}_{-3}$ & 2.506 & -2.38 \\
			ESO294-G010 & $26.41^{+0.02}_{-0.03}$ & $26.4^{+0.07}_{-0.07}$ & $26.47^{+0.01}_{-0.01}$ & $1918^{+25}_{-29}$ & $1905^{+61}_{-61}$ & $1913^{+18}_{-26}$ & $1979^{+16}_{-24}$ & 0.186 & -2.00 \\
			ESO410-G005 & $26.35^{+0.02}_{-0.03}$ & $26.33^{+0.07}_{-0.07}$ & $26.41^{+0.01}_{-0.01}$ & $1866^{+24}_{-28}$ & $1845^{+59}_{-59}$ & $1861^{+17}_{-26}$ & $1805^{+16}_{-23}$ & 0.316 & -2.00 \\
			Fornax & $20.68^{+0.02}_{-0.03}$ & $20.67^{+0.08}_{-0.08}$ & $20.75^{+0.00}_{-0.00}$ & $137^{+2}_{-2}$ & $136^{+5}_{-5}$ & $139^{+1}_{-2}$ & $765^{+0}_{-0}$ & 0.205 & -2.00 \\
			Grus I & $20.54^{+0.12}_{-0.12}$ & $20.51^{+0.1}_{-0.1}$ & $20.58^{+0.10}_{-0.10}$ & $129^{+7}_{-7}$ & $126^{+6}_{-6}$ & $124^{+7}_{-7}$ & $796^{+2}_{-2}$ & 0.224 & -2.5 \\
			Hercules & $20.58^{+0.07}_{-0.07}$ & $20.6^{+0.1}_{-0.1}$ & $20.58^{+0.07}_{-0.07}$ & $131^{+4}_{-4}$ & $132^{+6}_{-6}$ & $125^{+4}_{-4}$ & $819^{+2}_{-2}$ & 0.105 & -2.41 \\
			IC 1613 & $24.42^{+0.03}_{-0.03}$ & $24.4^{+0.07}_{-0.07}$ & $24.48^{+0.02}_{-0.01}$ & $768^{+12}_{-13}$ & $759^{+24}_{-24}$ & $768^{+11}_{-11}$ & $521^{+3}_{-3}$ & 0.352 & -2.00 \\
			Leo A & $24.40^{+0.06}_{-0.06}$ & $24.48^{+0.12}_{-0.12}$ & $24.46^{+0.05}_{-0.05}$ & $761^{+23}_{-24}$ & $787^{+43}_{-43}$ & $763^{+21}_{-21}$ & $1165^{+16}_{-16}$ & 0.529 & -2.00 \\
			Leo I & $21.90^{+0.03}_{-0.03}$ & $22.06^{+0.08}_{-0.08}$ & $21.97^{+0.01}_{-0.01}$ & $241^{+4}_{-4}$ & $258^{+10}_{-10}$ & $244^{+3}_{-3}$ & $906^{+2}_{-2}$ & 1.765 & -2.00 \\
			Leo IV & $20.99^{+0.09}_{-0.09}$ & $20.94^{+0.07}_{-0.07}$ & $20.98^{+0.10}_{-0.10}$ & $158^{+7}_{-7}$ & $154^{+5}_{-5}$ & $158^{+7}_{-7}$ & $895^{+5}_{-5}$ & 0.474 & -2.54 \\
			Leo P & $26.11^{+0.05}_{-0.06}$ & $26.04^{+0.21}_{-0.21}$ & $26.17^{+0.05}_{-0.05}$ & $1671^{+46}_{-47}$ & $1614^{+156}_{-156}$ & $1671^{+38}_{-46}$ & $2089^{+36}_{-43}$ & 0.345 & -2.00 \\
			Leo T & $23.13^{+0.16}_{-0.15}$ & $23.06^{+0.15}_{-0.15}$ & $23.19^{+0.15}_{-0.15}$ & $425^{+32}_{-31}$ & $409^{+28}_{-28}$ & $427^{+31}_{-29}$ & $988^{+21}_{-19}$ & 0.366 & -2.00 \\
			NGC 147$^{a}$ & $24.02^{+0.03}_{-0.03}$ & $24.15^{+0.09}_{-0.09}$ & $24.12^{+0.01}_{-0.01}$ & $637^{+9}_{-10}$ & $676^{+28}_{-28}$ & $641^{+9}_{-9}$ & $167^{+7}_{-7}$ & 1.361 & -2.00 \\
			NGC 185$^{a}$ & $23.74^{+0.02}_{-0.03}$ & $23.95^{+0.09}_{-0.09}$ & $23.85^{+0.00}_{-0.00}$ & $562^{+7}_{-8}$ & $617^{+26}_{-26}$ & $564^{+5}_{-8}$ & $231^{+5}_{-7}$ & 2.148 & -2.00 \\
			Phoenix I & $22.99^{+0.03}_{-0.03}$ & $23.09^{+0.1}_{-0.1}$ & $23.06^{+0.01}_{-0.01}$ & $397^{+6}_{-6}$ & $415^{+19}_{-19}$ & $396^{+5}_{-5}$ & $853^{+2}_{-2}$ & 0.936 & -2.00 \\
			Phoenix II & $19.85^{+0.16}_{-0.17}$ & $20.01^{+0.1}_{-0.1}$ & $19.90^{+0.15}_{-0.15}$ & $94^{+7}_{-7}$ & $100^{+5}_{-5}$ & $90^{+7}_{-7}$ & $792^{+2}_{-2}$ & 0.790 & -2.75 \\
			Sagittarius II & $18.93^{+0.16}_{-0.16}$ & $19.03^{+0.1}_{-0.1}$ & $18.94^{+0.17}_{-0.17}$ & $61^{+5}_{-5}$ & $64^{+3}_{-3}$ & $54^{+4}_{-4}$ & $781^{+1}_{-1}$ & 0.490 & -2.1 \\
			Sculptor & $19.56^{+0.02}_{-0.03}$ & $19.62^{+0.04}_{-0.04}$ & $19.62^{+0.00}_{-0.00}$ & $82^{+1}_{-1}$ & $84^{+2}_{-2}$ & $82^{+1}_{-1}$ & $759^{+0}_{-0}$ & 1.175 & -2.00 \\
			Segue II & $17.82^{+0.16}_{-0.16}$ & $17.82^{+0.15}_{-0.14}$ & $17.83^{+0.17}_{-0.18}$ & $37^{+3}_{-3}$ & $37^{+3}_{-2}$ & $42^{+3}_{-3}$ & $745^{+2}_{-2}$ & 0.007 & -2.00 \\
			Tucana & $24.73^{+0.02}_{-0.03}$ & $24.74^{+0.12}_{-0.12}$ & $24.8^{+0.01}_{-0.01}$ & $886^{+12}_{-13}$ & $887^{+49}_{-49}$ & $879^{+8}_{-12}$ & $1347^{+7}_{-10}$ & 0.017 & -2.00 \\
			Ursa Major I & $19.78^{+0.07}_{-0.07}$ & $19.94^{+0.13}_{-0.13}$ & $19.80^{+0.07}_{-0.07}$ & $91^{+3}_{-3}$ & $97^{+6}_{-6}$ & $95^{+3}_{-3}$ & $770^{+0}_{-0}$ & 1.004 & -2.18 \\
			Ursa Major II & $18.29^{+0.15}_{-0.15}$ & $17.7^{+0.13}_{-0.13}$ & $18.29^{+0.17}_{-0.17}$ & $46^{+3}_{-3}$ & $35^{+2}_{-2}$ & $51^{+3}_{-3}$ & $759^{+1}_{-1}$ & 2.922 & -2.47 \\
			\enddata
		\end{deluxetable*}
	\end{longrotatetable}
	
	\newpage
	
	\newpage
	
	\section{Discussion}
	\label{sec:discussion}
	
	\subsection{Comparison to literature RR Lyrae distances}
	
	\subsubsection{Broad trends}
	
\begin{figure}[t!]
		\centering
		\includegraphics[width = \columnwidth]{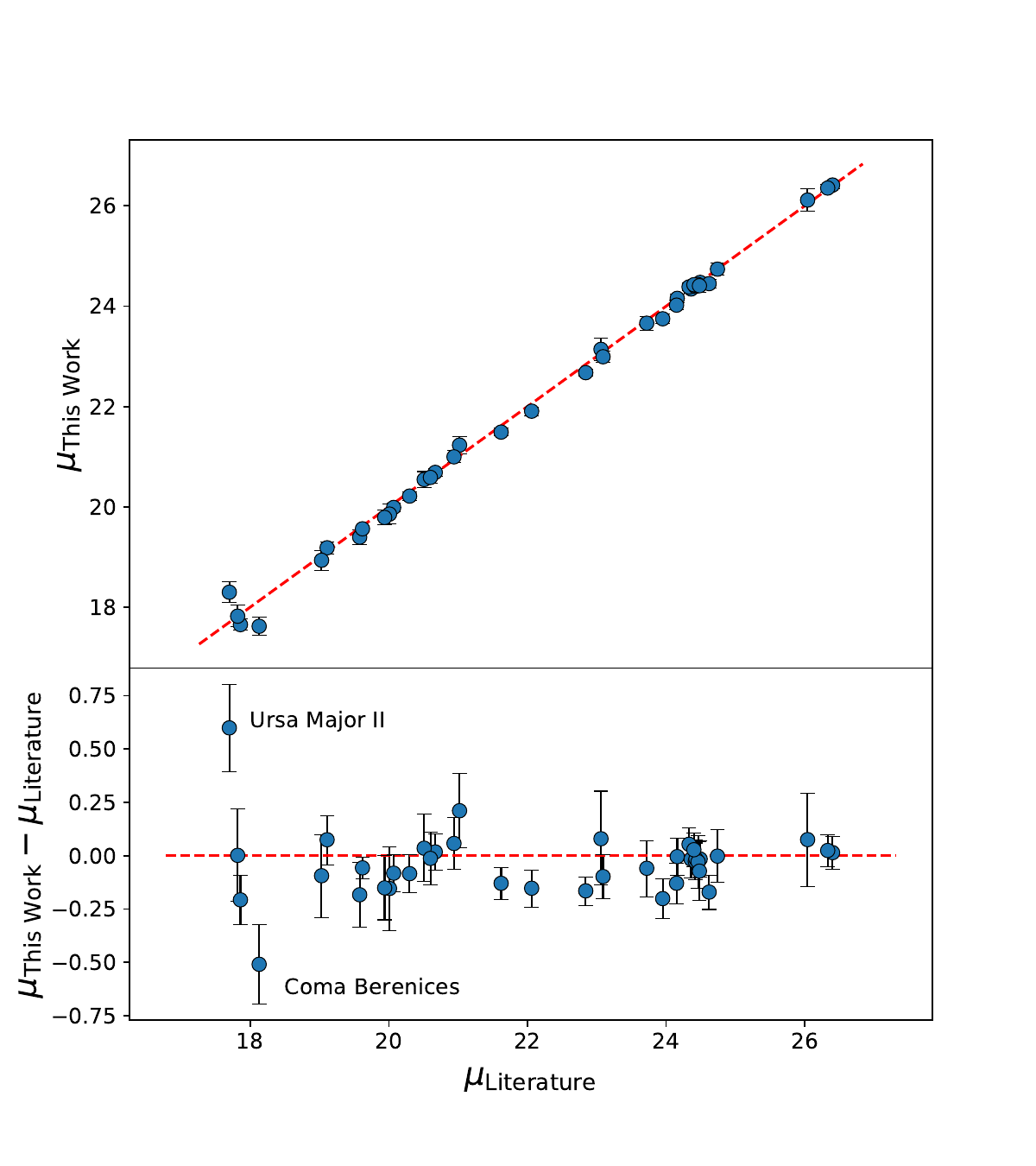}
		\caption{A comparison between our {\it Gaia} eDR3 based distance moduli and those from the literature (which are based on modeling the same RR Lyrae). The lower panel displays the residuals. The dashed lines reflect perfect agreement between the two distance determinations. Uncertainties reflect the quadrature sum of the error in our distance moduli and the literature distance moduli. We find generally good agreement with the literature. On average, our inferred distances are systematically closer than those in the literature by $\sim0.05$~mag.} 
		\label{fig:lit}
	\end{figure}

	Figure \ref{fig:lit} shows a comparison between our {\it Gaia}-anchored distances and literature distances to the same galaxies. The latter are taken from the paper presenting the RR Lyrae data we fit (se Table~\ref{tab:lgdata}). These literature values are  listed in Table \ref{tab:final}. Overall, we find that 27 galaxies ($\sim70$\%) in our sample agree within $1 \sigma$ with literature values, 4 galaxies ($\sim10$\%) agree within $(1 -1.5) \sigma$, and 3 galaxies ($\sim8$\%) agree with the literature to between $(1.5-2) \sigma$. The remaining 5 galaxies ($\sim13$\%) disagree at $> 2\sigma$.  We discuss the largest outliers individually in \S \ref{sec:outliers}.

	\subsubsection{Illustrative distance comparison: IC~1613}
	\label{sec:ic1613}
	
	To illustrate our distance determinations relative to the broader literature, we consider the case of IC~1613, a star-forming dwarf galaxy located at the edge of the LG.  IC~1613 has a low inclination angle and relatively little internal or foreground dust, making it particularly well-suited for distance determinations at low-metallicity.  According to NED, there have been over 100 distance measurements published for IC~1613.
	
	Figure \ref{fig:ic1613} shows Population~{\sc II} (i.e., TRGB and RR Lyrae) distances for IC~1613 that have been published over the last decade.  There is a scatter of $\sim0.15$~mag in both TRGB and RR Lyrae distances, with most older (i.e., pre-2014) estimates closer to $\mu\sim24.45$ and more recent estimates closer to $\mu\sim24.3$.
	
	We highlight two recent RR Lyrae distances to IC~1613.  The first is that of \citet{bernard_acs_2010}.  Using high cadence \textit{HST}-based imaging (which produce a very observed tight PW relation; see Figure \ref{fig:PLVI}), \citet{bernard_acs_2010} find a distance of $\mu=24.39\pm0.12$, assuming a mean metallicity of $\rm [Fe/H]=-1.6$ and the RR Lyrae period-luminosity relationship from \citet{clementini2003}. That metallicity assumption was motivated by a preliminary version of the star formation history (SFH) measured from IC~1613's deep color-magnitude diagram (CMD), eventually published in \citet{skillman2014}. 
	
	A more recent RR Lyrae distance was reported by \citet{hatt2017}, who find an average RR Lyrae-based distance for IC~1613 of $\mu = 24.28 \pm 0.04$.  This value is an average over 9 different RR Lyrae distances they determine in different permutations of bandpasses and RR Lyrae type (e.g., how RRab and RRc are treated).  They adopt multiband PWZs from the theoretical work of \citet{marconi2015} and assume a mean metallicity of $\rm [Fe/H]=-1.2$, which is taken from the dwarf galaxy luminosity-metallicity relationship \citep[e.g.,][]{kirby2013}.
	
	Our distance to IC~1613 is $\mu=24.42\pm0.03$.  We anchor our data with the \citet{Neeley2019} RR Lyrae data with updated {\it Gaia} eDR3 parallaxes (see Table \ref{tab:mwrrl}) and adopt a mean metallicity of $\rm [Fe/H]=-2.0$.  Though our selected metallicity may seem too low given the other studies, we argue that it is a reasonable choice.  Our adopted metallicity falls within the CMD-based age-metallicity relationship measured by \citet{skillman2014}.  For stars older than $\sim$10~Gyr, they find a mean metallicity of $\rm [Fe/H] = -1.75_{-0.15}^{+0.25}$.  This value was derived with Solar scaled stellar models; if the older population of IC~1613 are $\alpha$-enhanced, as they are in many dwarf galaxies \citep[e.g.,][]{Kirby2010}, the mean value of $\rm [Fe/H]$ could be up to 0.3~dex lower.  In comparison, the \citet{bernard_acs_2010} metallicity is slightly higher as it was adopted from an earlier version of the \citet{skillman2014} SFH.  Similarly, the specroscopic metallicity for IC~1613 is based on red giant branch stars \citep{kirby2013}, which are predominantly located in the central region of the galaxy.  Given IC~1613's continuous SFH \citep{skillman2014}, and population gradient \citet{skillman2014, weisz2014a}, it is likely that many of these RGB star correspond to ages younger than the RR Lyrae.  They would therefore be more metal-rich on average.
	
	To illustrate the effects of metallicity on the RR Lyrae distance to IC~1613, we recomputed our model assuming the spectroscopic metallicity of $\rm [Fe/H]=-1.2$.  The resulting distance is $\mu=24.28\pm0.03$, which is plotted in Figure \ref{fig:ic1613}.  It is in excellent agreement with that of \citet{hatt2017}.  As we discuss in \S \ref{sec:metallicity}, the adopted metallicity is an outstanding source of uncertainty in RR Lyrae-based distances.
	
	Finally, our distances are directly anchored to {\it Gaia} geometric parallaxes to metal-poor RR Lyrae.  As discussed in \citet{hatt2017}, this would be the preferred distance anchor, but the parallaxes were not available prior to publication of their paper.  \citet{bernard_acs_2010} anchor their distance to a previous study of RR Lyrae in the LMC.

	\subsubsection{The role of Adopted Metallicity}
	\label{sec:metallicity}
	As our analysis of IC~1613 highlights, the choice of mean RR Lyrae metallicity can influence the distance determination.  The amplitude of the effect can be considerable in some cases.  For example, for $V$-$I$ galaxies, the PWZ coefficients for the \textit{Gaia} anchored sample (Table~\ref{tab:mwrrl}) indicated a change in $\mu$ of $0.17$~mag for each $1~$dex change in mean metallicity.  While it may seem reasonable to assume that the adopted metallicity is known to better than 1~dex, the difference in our adopted metallicity for IC~1613 RR Lyrae is 0.8~dex smaller than \citet{hatt2017}.  
	
	The primary issue is that direct metallicities of RR Lyrae in many LG dwarf galaxies are challenging to measure due to their faintness.  To date, we are aware of only one paper in the literature that has spectroscopic metallicities of RR Lyrae in a LG dwarf galaxy \citep[Sculptor;][]{clementini2005}. Of the $\sim100$ RR Lyrae in Sculptor with spectroscopic metallicities, \citet{clementini2005} found that $\sim75$\% had [Fe/H] $< -1.7$ and only a single RR Lyrae had [Fe/H] $> -1.7$.  They conclude that the RR Lyrae population are most clearly associated with the ancient, metal-poor burst of star formation that created the blue HB.  Aside from this excellent study, in absence of RR Lyrae metallicities, other methods must be used. 
	
	The most obvious alternative is to adopt the mean metallicity from RGB spectra or the dwarf galaxy luminosity-metallicity relationship, which is based on RGB star spectra.  However, there are several selection effects that can complicate the mapping from RGB spectroscopic metallicities to RR Lyrae.  First, RR Lyrae are tracers of ancient populations, whereas in many cases, RGB stars are a mixed age population.  For galaxies that are purely ancient (e.g., ultra-faint dwarfs) the RGB-based metallicities are likely to be in good agreement with RR Lyrae metallicities.  However, in galaxies with more extended SFHs, the RGB metallicities may be biased toward higher values as they include younger, and presumably more metal-rich, stars.  This appears to be one of the issues in the case of IC~1613: the mean spectroscopic metallicity of $\rm [Fe/H]=-1.2$ reported by \citet{kirby2013} is likely to be representative of a mixed age population.  In contrast, the CMD-based age-metallicity from \citet{skillman2014} reports a mean metallicity of $\rm [Fe/H]=-1.75$ for stars older than $\sim10$~Gyr, a value that is more in line with the notion that RR~Lyrae in dwarf galaxies are ancient metal-poor stars.  Though deep CMD analysis can provide more age-appropriate metallicities, there are other issues such as the variation in absolute metallicity due to choice in stellar model (and their mapping to RR Lyrae) and the effect of possible $\alpha$-enhancements, which can lower the [Fe/H] by $\lesssim0.3$~dex for a fixed amount of total metals, $Z$.  
	
	Practical considerations may also affect the mapping of RGB star metallicities to RR Lyrae.  For example, maximizing slitmask efficiency may lead to an oversampling of RGB stars in the central region of a galaxy, thus leading to higher mean metallicities in the face of metallicity gradients, which appear to be common in dwarf galaxies \citep[e.g.,][]{held1999, saviane2000, tolstoy2004, deboer2012a, kacharov2017}

	Similarly, observational program design may adopt a fixed magnitude limit, which can lead to uneven age/metallicity sampling on the RGB \citep[e.g.,][]{Cole:2004ij}.  
	
	Given these challenges, and our prior that RR Lyrae in dwarf galaxies should be among the most metal-poor stars, we adopted a mean RR Lyrae metallicity of $\rm [Fe/H]=-2$ in all cases where the spectroscopic metallicity was more metal-rich than this value.  This is clearly a crude assumption as we would expect variations in the exact RR Lyrae metallicity based on a galaxy's true SFH.  However, we believe that this is a reasonable approximation in absence of direct RR Lyrae metallicities and potential biases in taking RGB star metallicities at face value. Direct spectroscopic metallicity measurements of RR Lyrae in nearby dwarfs provide a promising avenue to improve on this work.

	\subsection{Outliers Relative to the Literature}
	\label{sec:outliers}
	
	\subsubsection{Andromeda XIII}
	
	Based on \textit{HST}/WFPC2 RR Lyrae data, \citet{yang_hst/wfpc2_2012} report a distance of $\mu=24.62\pm0.05$. We find $\mu=24.44\pm0.06$ using the same data, a $\sim2$-$\sigma$ disagreement.  There are a few subtle differences in the analyses, which may reconcile the difference.  First, \citet{yang_hst/wfpc2_2012} adopt a metallicity of $[Fe/H] = -1.6$ based on the observed color of the RGB, whereas we assume $[Fe/H] = -2$.  Second,  \citet{yang_hst/wfpc2_2012} derive their own extinction, $A_V=0.341$, which is 0.12~mag larger than \citet{schlafly2011}.  We use the reddening-free PWZ.  Finally, they anchor their distances to the \citet{Chaboyer1999} RR Lyrae calibration, which is ultimately based on \textit{Hipparcos}.  As we discuss in \S \ref{sec:caretta}, this results in an average offset of $\sim0.05$~mag compared to \textit{Gaia} eDR3.
	
	Other literature distances to Andromeda~XIII are from the TRGB and range from $\mu=24.40^{+0.33}_{-0.49}$ \citep{conn2012} to $\mu=24.80^{+0.07}_{-0.42}$ \citep{collins2010} and are consistent with both our derived RR Lyrae distance and that of \citet{yang_hst/wfpc2_2012}.

	\begin{figure}[t!]
		\centering
		\includegraphics[width = \columnwidth]{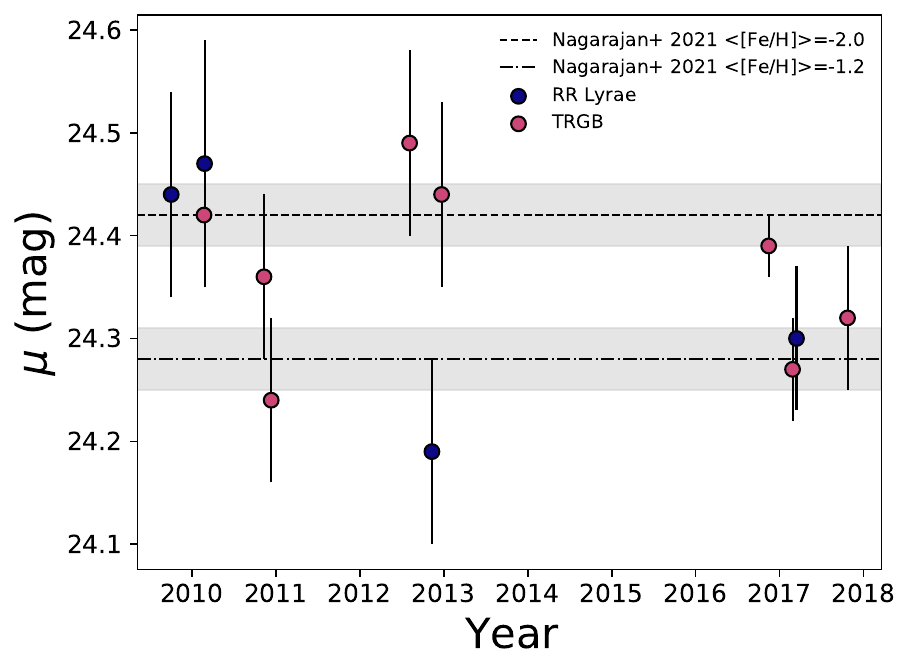}
		\caption{Published RR Lyrae and TRGB distances for IC~1613 since 2010.  We overplot our {\it Gaia}-anchored distance assuming a metallicity of $\rm [Fe/H]=-2.0$ (dashed line) and $\rm [Fe/H]=-1.2$ (dot-dashed line). As we discuss in \S~\ref{sec:ic1613} and \S~\ref{sec:metallicity}, unknown RR Lyrae metallicity (because RR Lyrae are generally too faint for direct spectroscopic metallicity determination at these distances) is among the largest sources of uncertainty in RR Lyrae-based galaxy distances.}
		\label{fig:ic1613}
	\end{figure}

	\subsubsection{Coma Berenices}
	
	Coma Berenices is an ultra-faint dwarf galaxy in the $V$-$I$ sample with a single RRab variable star \citep{musella_pulsating_2009}.  They report an RR Lyrae-based distance of $\mu=18.13 \pm 0.08$, while also noting that the RR Lyrae appears `over-luminous' by $\sim0.2$~mag compared to the mean HB magnitude.  This can explain a large part of the $\sim2.7$-$\sigma$ discrepancy with our distance $\mu=17.63 \pm 0.16$.  Coma~Berenices is too faint for an HB or TRGB distance determination.  \citet{brown2014} fit isochrones to deep HST-based CMDs and find $\mu=17.95 \pm 0.06$, though with an extinction that is $\sim3$ times larger than \citet{schlafly2011}.

	\subsubsection{Eridanus II}
	
	Using the same data, we find $\mu=22.67 \pm 0.04$ for Eridanus II, which is $\sim0.2$~mag lower than the distance of $\mu=22.84 \pm 0.05$ reported by  \citet{martinez-vazquez_monelli_cassisi_et_al_2021}. The two studies adopt the same metallicity ([Fe/H] $=-2.4$); \citealt{li2017}), though use very different methodologies to derive the distance.  Whereas we use the PWZ, \citet{martinez-vazquez_monelli_cassisi_et_al_2021} using the I-band PLZ along with a calibration ultimately tied to \textit{Hipparcos}.  It's unclear that this is sufficient to account for a 0.2 mag difference.
	
	Literature distances to Eridanus~II span a wide range that are consistent with both values. There are a set of larger distances ($\mu=22.8-22.9$; \citealt{koposov2015}, \citealt{crnojevic2016}, \citealt{gallart2021}) and smaller distances ($\mu=22.6-22.7$; \citealt{bechtol2015}, \citealt{simon2021}). At present, the primary reason for the tension between the two groups of distance determinations is unclear.

	\subsubsection{Ursa Major II}
	
	\citet{dallora_stellar_2012} report the detection of a single RRab star in Ursa Major II. They find $\mu=17.70 \pm 0.13$, whereas, with the same data, we find $\mu=18.29 \pm 0.15$.  In trying to reconcile the difference, we found we could not replicate the distance quoted in \citet{dallora_stellar_2012}. Specifically, they use the RR Lyrae calibration of \citet{clementini2003}:
	
	\begin{equation}
	\begin{split}
	M_V & = (0.214 \pm 0.047) \times (\rm [Fe/H] + 1.5) \\ & + (0.54 \pm 0.017 \pm 0.085)
	\end{split}
	\end{equation}
	
	for $\rm [Fe/H]=-2.47$.  This results in $M_V = 0.34$. Since the RRab variable has $m_V=18.39$, this leads to $\mu=17.95 \pm 0.10$ after accounting for an extinction of $E(B - V) = 0.096$ \citep{Schlegel1998}. The distance we derive using the data from \citet{dallora_stellar_2012} and calibration as stated in the paper is in much better agreement with our \textit{Gaia}-based distance.  Though we are not able to reproduce the literature RR Lyrae distance using their same data and calibration, we do note that the above \citet{clementini2003} calibration is only valid for [Fe/H]$ >-2.1$, whereas Ursa~Major~II is $\sim0.4$~dex more metal-poor.  Other literature distances to Ursa~Major~II are from fitting CMDs of the very shallow discovery SDSS data, or follow up data, and report a range of distances from $\mu=17.5 \pm 0.3$ to $\mu=17.90 \pm 0.23$ \citep[e.g.,][]{zucker2006, liu2008}.
	
	\begin{deluxetable}{ccc}
		\tablecaption{Theoretical Parameters for B-V and V-I PWZ Relations from \citet{marconi2015}. Note that these values are for fundamental mode pulsators. The PWZ relations take the form M = a + b $\log{P}$ + c [Fe/H].
			\label{tab:marconi}}
		\tablehead{\colhead{Parameter} &
			\colhead{Description} & \colhead{Theoretical Value}}
		\startdata
		$a_{BV}$ & Zero Point for B-V PWZ & -1.11 \\
		$b_{BV}$ & Period Slope for B-V PWZ & -2.67 \\
		$c_{BV}$ & Metallicity Slope for B-V PWZ & -0.02 \\
		$a_{VI}$ & Zero Point for V-I PWZ & -0.94 \\
		$b_{VI}$ & Period Slope for V-I PWZ & -2.43 \\
		$c_{VI}$ & Metallicity Slope for V-I PWZ & 0.15 \\
		\enddata
	\end{deluxetable}
	
	\subsection{Comparison to Theory}
	\label{sec:theory}
	
	In addition to our {\it Gaia} eDR3-based distances, we also compute distances to each galaxy using theoretical PWZ coefficients from \citet{marconi2015}, which are commonly employed in the literature.  Briefly, \citet{marconi2015} determine these coefficients by using a grid of nonlinear, time-dependent convective hydrodynamical models to predict dual band PW relations across the a broad metallicity range ($Z = 0.0001$ to $0.02$). 
	
	To anchor our distances to \citet{marconi2015}, we tightly constrain our PWZ priors around 
	the \citet{marconi2015} coefficients, using a normal distribution with a mean equal to the theoretical value and $\sigma=0.001$. We list the \citet{marconi2015} coefficients for the $B$-$V$ and $V$-$I$ PWZ relations in Table \ref{tab:marconi}.  With these fixed PWZ parameters, we run the entire sample, minus the anchor sample, through the procedure described in Section \ref{sec:method}.
	
	\begin{figure}[t!]
		\centering
		\includegraphics[width = \columnwidth]{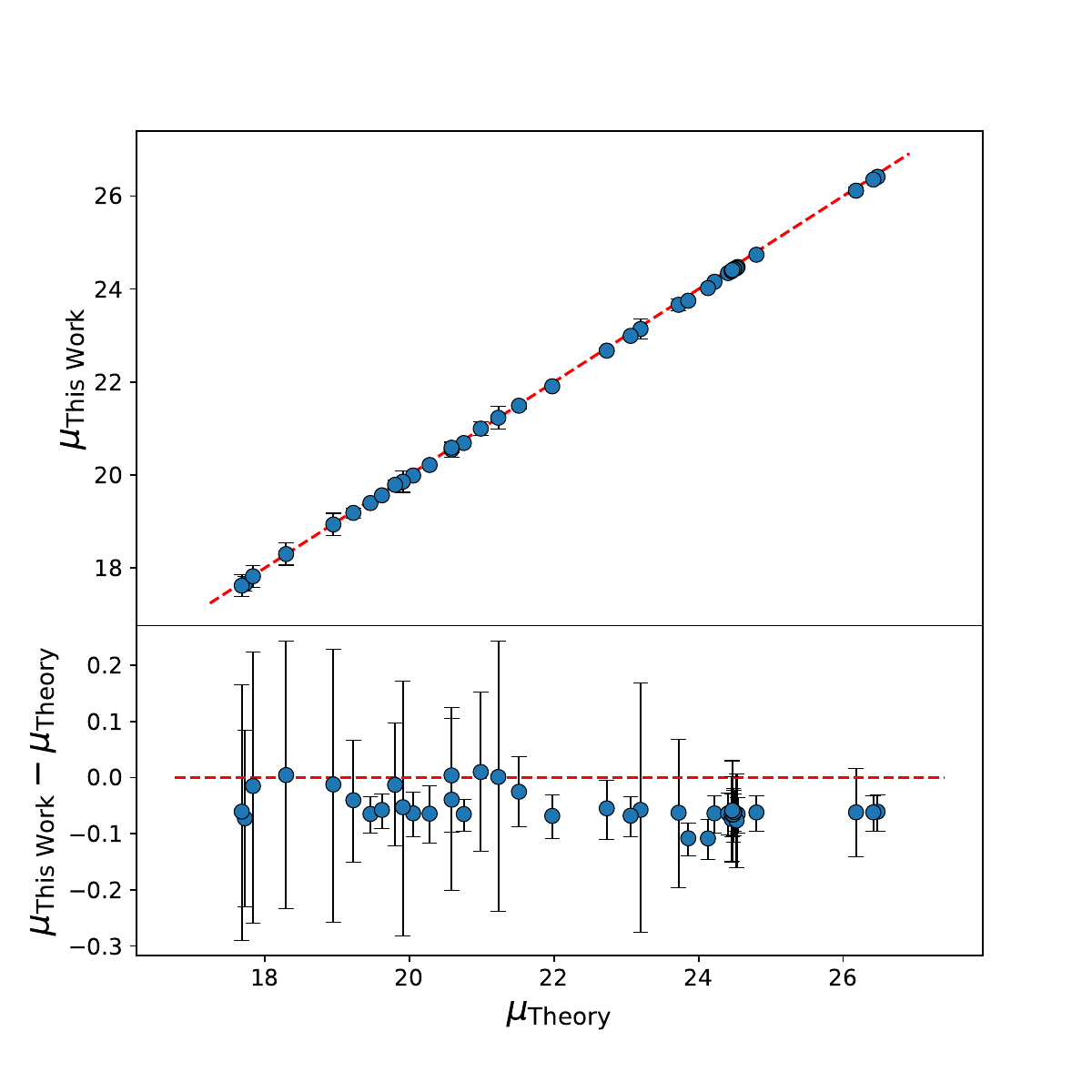}
		\caption{A comparison between our {\it Gaia} eDR3 based distance moduli and those based on PWZ parameters from \citet{marconi2015}. The lower panel displays the residuals. The dashed lines reflect perfect agreement between the two distance determinations. Uncertainties reflect the quadrature sum of the error in our distance moduli and the theoretical distance moduli. We find generally good agreement with \citet{marconi2015}. On average, our inferred distances for the V-I sample are systematically closer by $\sim0.06$~mag.}
		\label{fig:theory_comp}
	\end{figure}
	
	The resulting distances are listed in Table \ref{tab:final} as $\mu_{\rm Theory}$, while Figure \ref{fig:theory_comp} shows a comparison between our \textit{Gaia}-anchored distances and distances based on \citet{marconi2015} to the same galaxies.  The galaxy distances are in reasonable agreement; 60\% agree within 1-$\sigma$ and the remainder within 2-$\sigma$.  We do find that the \citet{marconi2015} PWZ leads to larger distances by $\sim0.06$~mag, on average, when compared to \textit{Gaia}-anchored distances.  We adopt the same metallicity for both anchors, thus; the disagreement is not due to metallicity. Instead, we find that the the value of $\alpha$, i.e., the parameter that controls the extinction correction for converting to Wesenheit magnitudes, contributes to this discrepancy.  \citet{marconi2015} use a value of $\alpha=1.38$ for V-I galaxies, compared to $\alpha=1.47$, which is adopted by \citet{Neeley2019}; these correspond to extinction curve ($R_V$) values that differ by $\sim0.05$.  If we use the \citet{marconi2015} anchor with the \citet{Neeley2019} value of $\alpha$, we find the distance discrepancy reduces to $\sim0.02$~mag, on average.  This is not an appropriate use of $\alpha$, as the \citet{marconi2015} PWZ coefficients are derived for a specific extinction curve that is different that what \citet{Neeley2019} adopted.  Nevertheless, it illustrates how uncertainty in the extinction is not totally mitigated by the use of Wesenheit magnitudes.  Overall, the level of agreement in the two distance anchors is generally quite good.

	\subsubsection{Comparison to Hipparcos-based Calibration}
	\label{sec:caretta}
	
	Prior to \textit{Gaia}, many Population II distance indicators were ultimately anchored to \textit{Hipparcos} geometric distances of stars in the Solar neighborhood.  None of these stars are RR Lyrae.  Instead, substantial effort went into using the very local metal-poor subdwarfs that did have \textit{Hipparcos} parallaxes to calibrate other Population II distance indicators as well as the distance to the LMC.  
	
	Among the most widely used of such calibrations is that of \citet{carretta2000}, in which various HB and RR Lyrae distance calibrations are presented.  These were ultimately used to also calibrate the widely used TRGB calibration presented in \citet{rizzi2007}.
	
	The RR Lyrae calibration recommended in \citet{carretta2000} is:
	
	\begin{equation}
	\label{eq:carreta}
	M_V(RR) = (0.18 \pm 0.09) (\rm [Fe/H] + 1.5) + (0.57 \pm 0.07)
	\end{equation}
	where $M_V(RR)$ is the (extinction-corrected) absolute $V$-band magnitude of an RR Lyrae and $\rm [Fe/H]$ is its metallicity. The relation does not include any dependence on pulsation period and can thus be regarded as predicting the ensemble mean magnitude of many RR Lyrae with a typical period distribution. 
	
	To assess the consistency of this calibration with our distance scale, we applied this \citet{carretta2000} relation to the Milky Way RR Lyrae in our anchor sample, which have spectroscopic [Fe/H] measurements and well-characterized mean magnitudes with modest extinction. 
	
\begin{figure}[t!]
		\centering
		\includegraphics[width = \columnwidth]{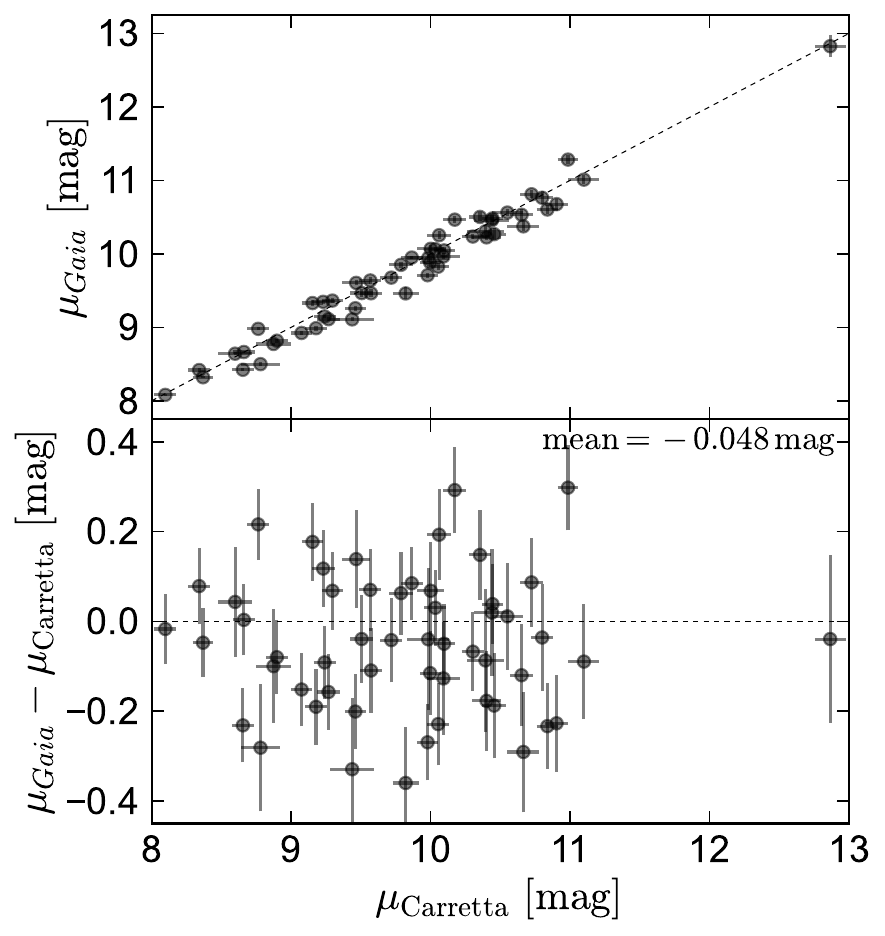}
		\caption{Comparison between the zero point corrected {\it Gaia} eDR3-based distance moduli for RR Lyrae in the Milky Way anchor sample (Figure~\ref{fig:gaia}) and predictions for the same stars based on the RR Lyrae absolute magnitude-metallicity relation from \citet{carretta2000}. The latter is based on a {\it Hipparcos}-anchored distance scale and has been widely used to anchor Population II distance indicators. On average, the \citet{carretta2000} relation predicts distance moduli that are 0.05 mag more distant than {\it Gaia} eDR3 parallaxes. This offset is comparable to between our inferred distances to dwarf galaxies and previous literature estimates (Figure~\ref{fig:lit}).}
		\label{fig:caretta}
	\end{figure}

	Figure~\ref{fig:caretta} shows the anchor sample distances predicted by \citet{carretta2000} versus the zeropoint-corrected \textit{Gaia} eDR3 geometric distances for our 36 RR Lyrae anchor stars. Distances anchored to the \citet{carretta2000} scale are, on average, $\sim$ 0.05 mag farther than those anchored to the {\it Gaia} eDR3 parallaxes. This value is consistent with the typical offset between the literature dwarf galaxy distances and what we infer using the \textit{Gaia} anchor sample (\S \ref{sec:results}).  There are many consequences of this shift down the road (e.g., the luminosity of the TRGB), which are beyond the scope of the present paper.
	
	To assess the consistency of this calibration with our distance scale, we applied it to the Milky Way RR Lyrae in our anchor sample, which have spectroscopic [Fe/H] measurements and well-characterized mean magnitudes with modest extinction. The results are shown in Figure~\ref{fig:caretta}, which compares the distance moduli to RR Lyrae in the anchor sample predicted by Equation~\ref{eq:carreta} to those calculated from the (zeropoint-corrected) {\it Gaia} eDR3 parallaxes. We find that the distances anchored to the \citet{carretta2000} scale are $\sim$ 0.05 mag farther than those anchored to the {\it Gaia} eDR3 parallaxes. This is consistent with our inferred dwarf galaxy distances, which are on average $\sim0.05$ closer than those in the literature.

	\subsection{Effects of parallax systematics}
	\label{sec:discussion_parallax}
	One of the significant methodological improvements of this work is our simultaneous modeling of RR Lyrae in dwarf galaxies and Milky Way RR Lyrae with well-determined parallaxes from {\it Gaia} eDR3. The improvement in the parallaxes of stars in this  ``anchor sample'' compared to {\it Gaia} DR2 is obvious (e.g. Figure~\ref{fig:gaia}). As a result of improved formal parallax uncertainties (factor of 2.1 on average) and improved treatment of systematics, the metallicity-dependence of the PWZ is obvious in the anchor sample with eDR3 data. This was not the case with DR2 data. 
	
	Nevertheless, some systematic uncertainties in parallaxes persist. Perhaps most importantly, the {\it Gaia} parallax zeropoint is somewhat uncertain, and likely depends on source properties. In Appendix~\ref{sec:zpt}, we investigate the sensitivity of our model to the zeropoint correction. Within the Milky Way anchor sample, the correction from \citet{Lindegren2020zpt} typically shifts the distance modulus by -0.03 mag (i.e., bringing the sources closer). This magnitude of this shift is similar to the typical uncertainty in distance modulus due to parallax errors, but it is systematic, shifting all sources in the same direction. There is little doubt that applying the \citet{Lindegren2020zpt} zerpoint correction results in more reliable parallaxes than using the uncorrected values, but there is some evidence that systematics remain at the 0.01 mas level \citep[e.g.,][]{Elbadry2021, Zinn2021, Riess2021},  corresponding to distance modulus errors of order 0.01 mag. Further reduction of parallax systematics with future {\it Gaia} data releases is one avenue along which our model could be improved.

	\subsection{Folding new distances into this framework}
	\label{sec:framework}
	
	Our framework can readily incorporate RR Lyrae from new dwarf galaxies in two different ways.  The most straightforward way to place new RR Lyrae distances on our scale, and that of \textit{Gaia} eDR3, is using the derived PWZ parameters in Table \ref{tab:mwrrl}, along with the form of Equation 7, to determine the distance modulus to that galaxy. As written, this method requires apparent magnitudes for each RRL in two different visual bands (either $V$-$I$ or $B$-$V$), along with knowledge of each variable's period and metallicity. As is the case for most dwarf galaxies, RR Lyrae metallicities are usually not directly measured.  In such cases, we recommend, at least for the sake of consistency, adopting our approach as described in Section \ref{sec:method}.
	
	A second approach is to run new data through the full model.  This has the benefit of including all covariances in the resulting uncertainties.  We have made our code, along with documentation for using it, publicly available in a Github repository\footnote{\url{https://github.com/pranav-nagarajan/Mapping-Local-Group}}.

	\subsection{Recommendations for improved RR Lyrae distances to dwarf galaxies}
	\label{sec:recommend}
	
	By virtue of using heterogeneous data from the literature, our model, and results, are only as good as the fidelity of the input data.  We make several recommendations for improving the data reliability going forward.  
	
	The first is to ensure adequate light curve sampling.  Our model assumes that errors in period are negligibly small.  This appears to be the case, as most of the existing light curves are very well-sampled.  However, this may not always be the case in future studies and our approach would have to be modified (i.e., the likelihood function would need to handle uncertainties in period and magnitude).  This problem has a known solution, but it can be non-trivial to implement \citep[e.g.,][]{hogg2010}.  
	
	A second, and related, issue is ensuring small uncertainties on the mean magnitudes.  The color term in the Wesenheit magnitude conversion amplifies the photometric uncertainties. Thus, if the mean magnitudes have uncertainties of more than a few hundredths of a magnitude, the constraints on the model free parameters (i.e., PWZ coefficients, distance) will be broad.
	
	Third, is the use of a common filter system.  The need to transform filters introduces another dimension of uncertainty into our fits. As discussed in seminal papers on filter transformations \citep[e.g.,][]{sirianni2005, saha2011}, filter transformations should be avoided whenever possible. Thus, either a common filter systems needs to be used for future RR Lyrae data collection or more investment needs to be made in establishing a common set of filter transformations among the many different systems (e.g., \textit{HST}, DES, SDSS, Pan-STARSS, Rubin).  One workaround would be to observe the \textit{Gaia} RR Lyrae anchor sample in all of the filter sets commonly used, with suitable cadence to build light curves.  
	
	Finally, as discussed in \S \ref{sec:metallicity}, the PWZ relation relies on knowledge of the mean metallicity of the observed RR Lyrae. However, due a paucity of spectra of RR Lyrae outside a few MW satellites, RR Lyrae metallicities are unknown.  The adoption of ``reasonable'' metallicities in the literature can affect the distance by $\sim0.1$~mag, which is often in excess of other sources of uncertainty.  Improved calibrations of RR Lyrae metallicities, and their relationships to brighter proxies (e.g., RGB color, RGB metallicity) would be a substantial step in mitigating this source of uncertainty.

	\section{Summary}
	\label{sec:conclusions}
	
	The resolved stellar populations of nearby galaxies provide multiple `gold standard' distance indicators.  However, to date, virtually all distances are measured for single or small sets of galaxies.  The result is a large degree of heterogeneity in distance determinations and associated uncertainties, making it hard to place galaxy distances on the `same scale' needed for any ensemble study of galaxies (e.g., star formation or orbital histories).
	
	To remedy this situation, in this work, we present uniformly determined RR Lyrae-based distances to 39 galaxies in and around the Local Group. We employ a Bayesian hierarchical model and fit the Period-Wesenheit magnitude-Metallicity (PWZ) relation using published periods and apparent magnitudes of RRab stars. We anchor the PWZ relationship using a sample of metal-poor Milky Way RR Lyrae with published light curves, spectroscopic metallicities, and \textit{Gaia} eDR3 parallaxes, most of which were collected in \citet{Neeley2019}. The main findings from this paper are: 
	\begin{itemize}
		
		\item  As shown in Figure \ref{fig:parallaxes} and Table \ref{tab:final}, $\sim70$\% of our derived distances agree with the literature distances that use the same data to within 1-$\sigma$. On average, our distances are $\sim0.05$~mag closer than the literature values.  We believe this is due to being anchored to the updated \textit{Gaia} eDR3 RR Lyrae parallaxes, which place Galactic RR Lyrae closer by a similar amount. 
		
		\item Our distance uncertainties have been uniformly derived.  Our distances are also 2-3 times more precise than literature distances (e.g., Table \ref{tab:final}).  This is in large part due to the simultaneous modeling of galaxy distances, which accounts for covariant uncertainties.  
		
		\item  $\sim10$\% of galaxies are classified as outliers ($> 2 \sigma$ disagreement with literature values; e.g., Figure \ref{fig:lit}). We discuss these in \S \ref{sec:outliers}.  In most cases, we find anomalies in the literature distances (e.g., poor photometry, unusual zero point calibration).  In general, a wider survey of the literature reveals that our distances fall within the range of nearby dwarf galaxy distances from the last decade.
		
		\item The two main source of uncertainty in our distances are the unknown RR Lyrae metallicities (Figure \ref{fig:ic1613} and \S \ref{sec:metallicity}) in most dwarf galaxies and the heterogeneity in published RR Lyrae photometry for nearby galaxies (\S \ref{sec:recommend}).  Given the lack of direct RR Lyrae metallicities in dwarf galaxies and the unfortunate lack of photometric uncertainties reported for many RR Lyrae, we adopt literature-motivated assumptions for the metallicities and photometric uncertainties.  
		
		\item We apply the \textit{Hipparcos}-based distance calibration from \citet{carretta2000} to our \textit{Gaia} eDR3 anchor sample and find it over-predicts the \textit{Gaia}-based geometric distance by 0.05~mag on average (Figure \ref{fig:caretta} and \S \ref{sec:caretta}).
		
		\item We run our model using a PWZ anchor based on theoretical models of RR Lyrae presented by \citet{marconi2015}.  In Figure \ref{fig:theory_comp}, we show that the resulting distance estimates to be $\sim0.06$~mag larger, on average, when compared to our \textit{Gaia}-anchored distances. We discuss how this difference may be related to the adoption of slightly different extinction curves (\S \ref{sec:theory}).
		
		\item We conclude by providing code infrastructure and recommendations for including new RR Lyrae distances into our distance framework (\S \ref{sec:framework}) and by discussing data needed for improved RR Lyrae distances (\S \ref{sec:recommend}).
		
	\end{itemize}
	
	\begin{acknowledgments}
		The authors thank the anonymous referee for a constructive and helpful report. The authors also thank Alessandro Savino for useful discussion about RR Lyrae distances and Marcella Marconi and Roberto Molinaro for helpful comments. DRW acknowledges support from HST-GO-15476, HST-GO-15901, HST-GO-15902, HST-AR-16159, HST-GO-16226, and HST-AR-16632 from the Space Telescope Science Institute, which is operated by AURA, Inc., under NASA contract NAS5-26555. KE acknowledges support from an NSF graduate research fellowship.
		
	\end{acknowledgments}
	
	%
	
	
	\software{astropy \citep{2013A&A...558A..33A}}
	
	
	
	\appendix
	
	\section{Filter Transformations}
	\label{sec:filtertransforms}
	
	The \textit{Gaia} anchor sample has data in Johnson (or UBVRI) photometric system, whereas the dwarf galaxies have data in a variety of photometric systems (Table \ref{tab:lgdata}).
	
	To convert HST magnitudes to UBVRI magnitudes, we inverted the transformations given by \citet{saha2011}. For instance, given magnitudes in the F606W and F814W filters:
	
	\begin{equation}
	V - I = \frac{1}{0.812} ((F606W - F814W) + 0.930)
	\end{equation}
	\begin{equation}
	I = F814W + 25.480 + 0.024 * (V - I)
	\end{equation}
	
	To convert DES magnitudes to UBVRI magnitudes, we first convert to the SDSS photometric system by inverting the transformations given by \citet{DrlicaWagner2018}.  Then, we use the empirical transformations given by \citet{jordi2006} for metal-poor stars. To illustrate, given mean magnitudes in the $ugriz$ filters,
	
	\begin{equation}
	(g - r)_{SDSS} = \frac{1}{0.998}((g - r)_{DES} + 0.01)
	\end{equation}
	\begin{equation}
	(i - z)_{SDSS} = \frac{1}{0.830}((i - z)_{DES} - 0.01)
	\end{equation}
	\begin{equation}
	\begin{split}
	(g - i)_{SDSS} &= (g - i)_{DES} + 0.104 * (g - r)_{SDSS} \\ &- 0.256 * (i - z)_{SDSS} + 0.01
	\end{split}
	\end{equation}
	
	\begin{equation}
	\begin{split}
	& V - I = \\ &
	\begin{cases}
	0.674(g - i)_{SDSS} + 0.406, & \text{if } (g - i)_{SDSS} \leq 2.1\\
	0.674(g - i)_{SDSS}, & \text{otherwise}
	\end{cases} 
	\end{split}
	\end{equation}
	
	\begin{equation}
	V - g_{SDSS} = 0.021 - 0.569(g - r)_{SDSS}
	\end{equation}
	\begin{equation}
	g_{SDSS} = g_{DES} + 0.104(g - r)_{SDSS} - 0.01
	\end{equation}
	\begin{equation}
	I = (V - g_{SDSS}) + g_{SDSS} - (V - I)
	\end{equation}
	
	After deriving the $I$ magnitude and $V - I$ color for each RR Lyrae star, we can then compute the $V, I$ Wesenheit magnitude for each star as described in \S \ref{sec:wesenheit}.
	
	\section{Period-Wesenheit Relationships for Full Galaxy Sample}
	\label{sec:all_galaxies}
	
\begin{figure*}[h!]
		\centering
		\includegraphics[width = 0.9\textwidth]{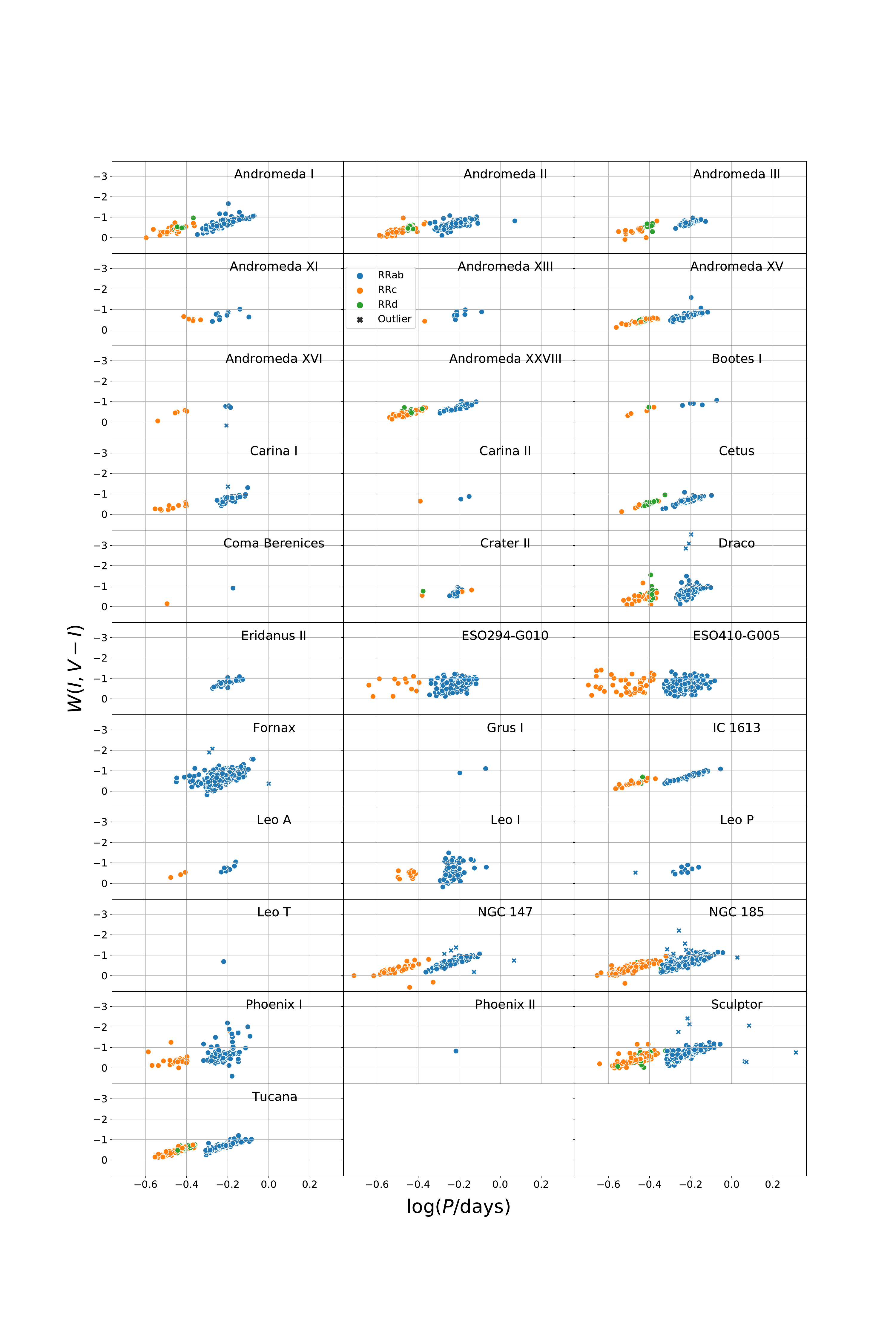}
		\caption{Observed period-Wesenheit magnitude relations for $V$-$I$ LG Galaxies.}
		\label{fig:PLVI}
	\end{figure*}

\begin{figure*}[h!]
		\centering
		\includegraphics[width = \textwidth]{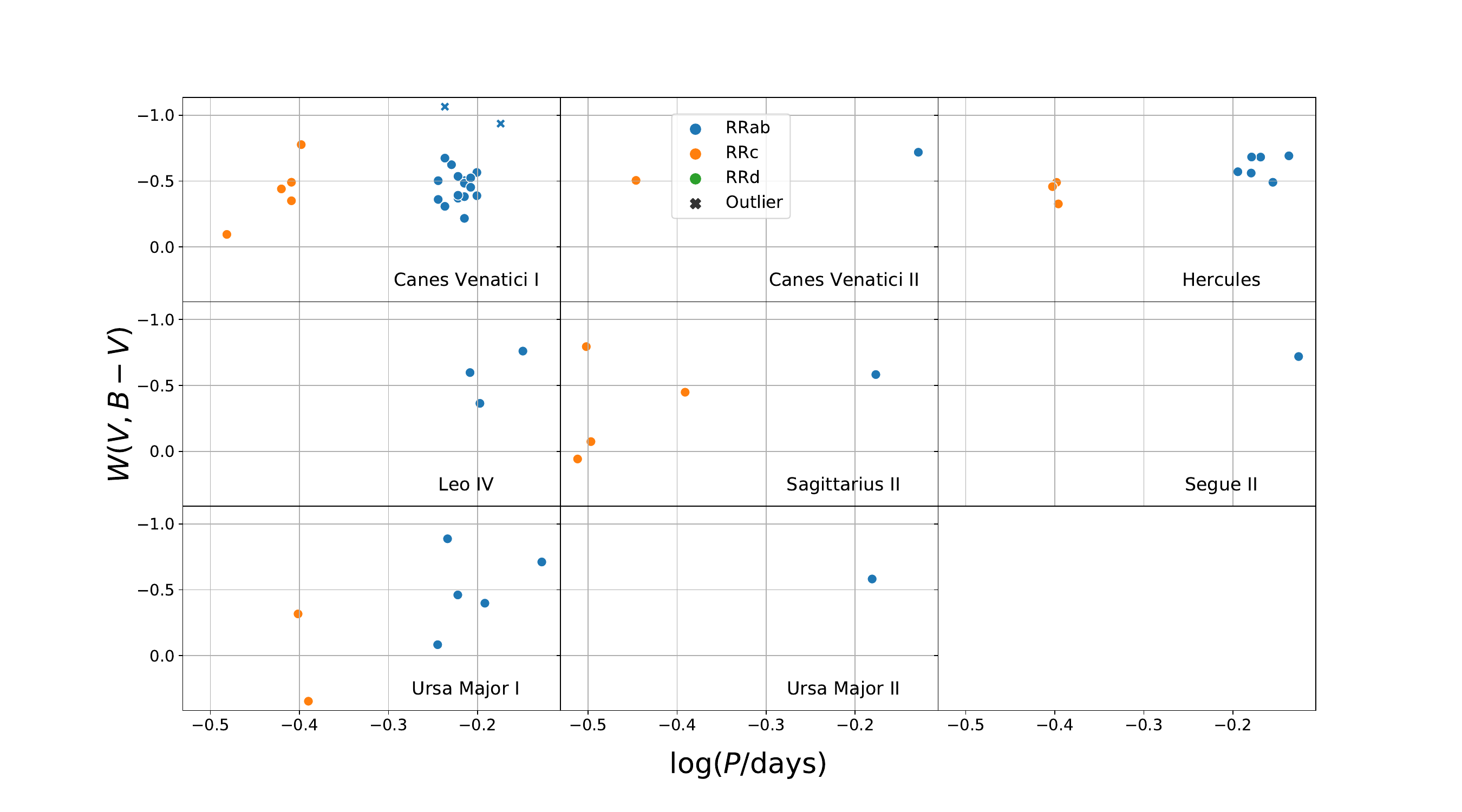}
		\caption{Observed period-Wesenheit magnitude relations for $B$-$V$ LG Galaxies.}
		\label{fig:PLBV}
	\end{figure*}

	Figures \ref{fig:PLVI} and \ref{fig:PLBV} show show the Period-Wesenheit relationships for the full samples of V-I and B-V galaxies, respectively.  A detailed description of the plot layout and discussion of general trends is presented in \S \ref{sec:data}. 
	
	\clearpage

	\section{Gaia parallax systematics}
	\label{sec:zpt}
	
\begin{figure*}
		\centering
		\includegraphics[width = \textwidth]{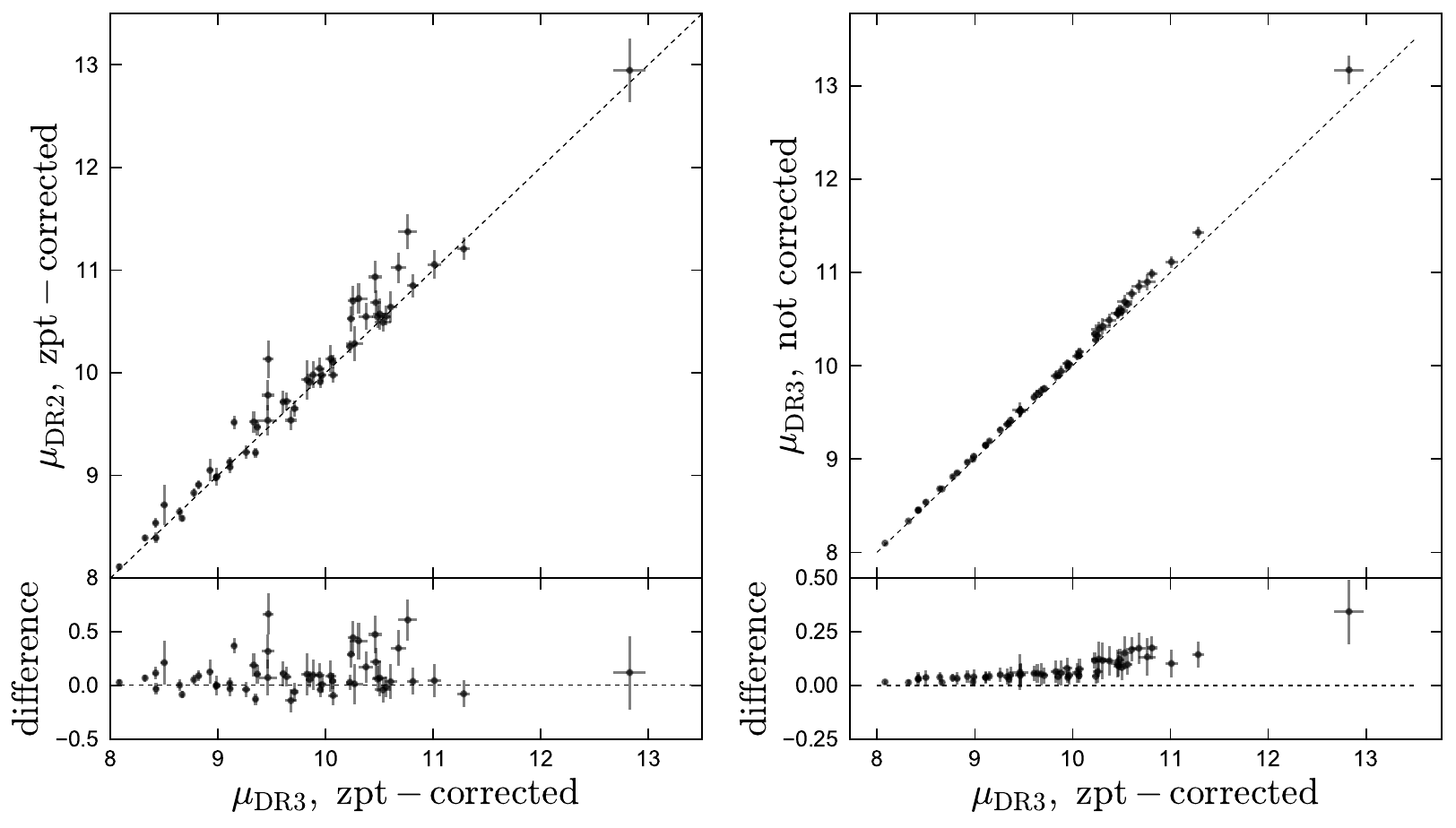}
		\caption{Effects of switching from DR2 to eDR3 parallaxes (left) and applying a parallax zeropoint correction (right) on the distance moduli of RR Lyrae in the Milky Way  sample. Note that the scales in the lower panels are different in the right and left columns. On average, using eDR3 parallaxes shifts distance moduli closer by 0.05 mag compared to DR2. The parallax zeropoint correction also decreases the inferred distance moduli, by a factor that is larger for more distant stars.}
		\label{fig:parallaxes}
	\end{figure*}

	Figure~\ref{fig:parallaxes} explores the effects of the using parallaxes from {\it Gaia} eDR3 vs. DR2, and of correcting the reported parallaxes for a nonzero parallax zeropoint. The left panel compares the DR2 and eDR3 parallaxes of the 54 RR Lyrae in the \citet{Neeley2017} sample. The sample contains 36 RR Lyrae of type RRab, which form our calibration sample, and 18 of type RRc/RRd. For the DR2 parallaxes, we apply a global parallax zeropoint correction of 30 $\mu$as \citep{lindgren2018}. For the eDR3 parallaxes, we apply the correction constructed by \citet{Lindegren2020zpt}, which depends on the color, magnitude, and sky position of the source. We calculate distance modulus as $\mu=-5\log\left(\varpi/100\right)$, where $\varpi$ represents the zeropoint-corrected parallax in mas. On average, the corrected RR Lyrae parallaxes are somewhat larger in eDR3 data (median difference of 0.023 mas), so the distance moduli are closer (median difference of 0.05 mag) in eDR3. As a result, using eDR3 parallaxes in the calibration sample reduces our inferred dwarf galaxy distances.
	
	The right panel compares distance moduli calculated from from the raw eDR3 parallaxes to those calculate from zeropoint-corrected parallaxes. Although the adopted zeropoint is different from each star, it is always negative, meaning that zeropoint-corrected parallaxes yield closer distances than the raw parallaxes. The effect of the zeropoint correction is largest for the most distant objects, since a fixed parallax offset translates to a larger distance offset at larger distances. The median offset in distance modulus between values calculated with and without the zeropoint correction is 0.026 mag. 
	
	\clearpage
	
	\section{Data for Milky Way RR Lyrae}
	\label{sec:mwrrlyrae}
	
	Table \ref{tab:mwrrl_summary} lists the properties of the MW RR Lyrae anchor sample used in our analysis.  All quantities, except for the parallax, are taken from \citet{Neeley2019}.  The parallax is from cross-matching these stars to \textit{Gaia} eDR3. We have applied the zero point correction from \citet{Lindegren2020zpt} and the inflated error bars from \citet{Elbadry2021} to these parallaxes.
	
	\startlongtable
	\begin{deluxetable*}{ccccccccc}
		\tablecaption{Summary of Milky Way RR Lyrae. (1) Name, (2) RR Lyrae class, (3) Period of pulsation in days, (4) Parallax from {\it Gaia} eDR3 in milliarcseconds, (5) Mean $B$ magnitude, (6) Mean $V$ magnitude, (7) Mean $I$ magnitude, (8) Spectroscopic Metallicity, (9) Extinction}
		\label{tab:mwrrl_summary}
		\tablehead{\colhead{Star} &
			\colhead{Class} & \colhead{Period}  & \colhead{Parallax} & \colhead{$\langle$B$\rangle$} & \colhead{$\langle$V$\rangle$} & \colhead{$\langle$I$\rangle$} & \colhead{[Fe/H]} & \colhead{E(B - V)} \\ 
			\colhead{} &
			\colhead{} & \colhead{(Days)}  & \colhead{(mas)} & \colhead{(mag)} & \colhead{(mag)} & \colhead{(mag)} & \colhead{(dex)} \\
			\colhead{(1)} &
			\colhead{(2)} & \colhead{(3)}  & \colhead{(4)} & \colhead{(5)} & \colhead{(6)} & \colhead{(7)} & \colhead{(8)} & \colhead{(9)}}
		\startdata
		AB UMa & RRab & 0.600 & 1.001 $\pm$ 0.021 & 11.359 $\pm$ 0.009 & 10.912 $\pm$ 0.009 & 10.342 $\pm$ 0.009 & -0.49 & 0.022 \\
		AE Boo & RRc & 0.315 & 1.128 $\pm$ 0.021 & 10.887 $\pm$ 0.009 & 10.64 $\pm$ 0.009 & 10.254 $\pm$ 0.009 & -1.39 & 0.023 \\
		AM Tuc & RRc & 0.406 & 0.553 $\pm$ 0.016 & 11.918 $\pm$ 0.006 & 11.626 $\pm$ 0.006 & 11.187 $\pm$ 0.006 & -1.49 & 0.023 \\
		AN Ser & RRab & 0.522 & 1.011 $\pm$ 0.024 & 11.321 $\pm$ 0.008 & 10.935 $\pm$ 0.008 & 10.446 $\pm$ 0.008 & -0.07 & 0.040 \\
		AP Ser & RRc & 0.341 & 0.800 $\pm$ 0.019 & 11.368 $\pm$ 0.008 & 11.129 $\pm$ 0.008 & 10.765 $\pm$ 0.008 & -1.58 & 0.042 \\
		AV Peg & RRab & 0.390 & 1.492 $\pm$ 0.019 & 10.862 $\pm$ 0.008 & 10.470 $\pm$ 0.008 & 9.971 $\pm$ 0.008 & -0.08 & 0.067 \\
		BB Pup & RRab & 0.481 & 0.628 $\pm$ 0.022 & 12.585 $\pm$ 0.006 & 12.159 $\pm$ 0.006 & 11.602 $\pm$ 0.006 & -0.60 & 0.105 \\
		BH Peg & RRab & 0.641 & 1.181 $\pm$ 0.025 & 10.899 $\pm$ 0.005 & 10.425 $\pm$ 0.005 & 9.8129 $\pm$ 0.005 & -1.22 & 0.077 \\
		BX Leo & RRc & 0.363 & 0.732 $\pm$ 0.026 & 11.831 $\pm$ 0.006 & 11.584 $\pm$ 0.006 & 11.21 $\pm$ 0.006 & -1.28 & 0.023 \\
		CS Eri & RRc & 0.311 & 2.151 $\pm$ 0.017 & 9.244 $\pm$ 0.006 & 9.01 $\pm$ 0.006 & 8.657 $\pm$ 0.006 & -1.41 & 0.018 \\
		CU Com & RRd & 0.406 & 0.272 $\pm$ 0.019 & 13.654 $\pm$ 0.009 & 13.345 $\pm$ 0.009 & 12.873 $\pm$ 0.009 & -2.38 & 0.023 \\
		DH Peg & RRc & 0.256 & 2.067 $\pm$ 0.023 & 9.8 $\pm$ 0.01 & 9.52 $\pm$ 0.01 & 9.113 $\pm$ 0.01 & -0.92 & 0.080 \\
		DX Del & RRab & 0.473 & 1.743 $\pm$ 0.016 & 10.359 $\pm$ 0.006 & 9.927 $\pm$ 0.006 & 9.367 $\pm$ 0.006 & -0.39 & 0.092 \\
		HK Pup & RRab & 0.734 & 0.806 $\pm$ 0.019 & 11.761 $\pm$ 0.005 & 11.312 $\pm$ 0.005 & 10.707 $\pm$ 0.005 & -1.11 & 0.160 \\
		MT Tel & RRc & 0.317 & 2.055 $\pm$ 0.033 & 9.1879 $\pm$ 0.006 & 8.9660 $\pm$ 0.006 & 8.6079 $\pm$ 0.006 & -1.85 & 0.038 \\
		RR Cet & RRab & 0.553 & 1.626 $\pm$ 0.023 & 10.075 $\pm$ 0.006 & 9.7229 $\pm$ 0.006 & 9.219 $\pm$ 0.006 & -1.45 & 0.022 \\
		RR Gem & RRab & 0.397 & 0.868 $\pm$ 0.036 & 11.689 $\pm$ 0.011 & 11.349 $\pm$ 0.011 & 10.874 $\pm$ 0.011 & -0.29 & 0.054 \\
		RR Leo & RRab & 0.452 & 1.083 $\pm$ 0.027 & 11.0 $\pm$ 0.006 & 10.716 $\pm$ 0.006 & 10.28 $\pm$ 0.006 & -1.60 & 0.037 \\
		RU Psc & RRc & 0.390 & 1.278 $\pm$ 0.032 & 10.458 $\pm$ 0.004 & 10.162 $\pm$ 0.004 & 9.729 $\pm$ 0.004 & -1.75 & 0.043 \\
		RU Scl & RRab & 0.493 & 1.265 $\pm$ 0.035 & 10.555 $\pm$ 0.004 & 10.238 $\pm$ 0.004 & 9.805 $\pm$ 0.004 & -1.27 & 0.018 \\
		RV CrB & RRc & 0.332 & 0.688 $\pm$ 0.018 & 11.618 $\pm$ 0.008 & 11.383 $\pm$ 0.006 & 11.012 $\pm$ 0.006 & -1.69 & 0.039 \\
		RV Oct & RRab & 0.571 & 1.008 $\pm$ 0.013 & 11.386 $\pm$ 0.006 & 10.953 $\pm$ 0.006 & 10.335 $\pm$ 0.006 & -1.71 & 0.180 \\
		RV UMa & RRab & 0.468 & 0.955 $\pm$ 0.015 & 10.979 $\pm$ 0.021 & 10.710 $\pm$ 0.021 & 10.335 $\pm$ 0.021 & -1.20 & 0.018 \\
		RX Eri & RRab & 0.587 & 1.707 $\pm$ 0.025 & 10.083 $\pm$ 0.003 & 9.675 $\pm$ 0.003 & 9.12 $\pm$ 0.003 & -1.33 & 0.058 \\
		RZ Cep & RRc & 0.309 & 2.404 $\pm$ 0.013 & 9.907 $\pm$ 0.02 & 9.396 $\pm$ 0.011 & 8.747 $\pm$ 0.013 & -1.77 & 0.250 \\
		ST Boo & RRab & 0.622 & 0.778 $\pm$ 0.023 & 11.265 $\pm$ 0.012 & 10.94 $\pm$ 0.012 & 10.484 $\pm$ 0.012 & -1.76 & 0.021 \\
		ST CVn & RRc & 0.329 & 0.782 $\pm$ 0.029 & 11.591 $\pm$ 0.01 & 11.337 $\pm$ 0.01 & 10.949 $\pm$ 0.01 & -1.07 & 0.012 \\
		SU Dra & RRab & 0.660 & 1.335 $\pm$ 0.016 & 10.124 $\pm$ 0.006 & 9.781 $\pm$ 0.006 & 9.286 $\pm$ 0.006 & -1.80 & 0.010 \\
		SV Eri & RRab & 0.714 & 1.360 $\pm$ 0.026 & 10.357 $\pm$ 0.004 & 9.949 $\pm$ 0.004 & 9.379 $\pm$ 0.004 & -1.70 & 0.085 \\
		SV Hya & RRab & 0.479 & 1.145 $\pm$ 0.030 & 10.849 $\pm$ 0.006 & 10.538 $\pm$ 0.006 & 10.059 $\pm$ 0.006 & -1.50 & 0.080 \\
		SV Scl & RRc & 0.377 & 0.703 $\pm$ 0.030 & 11.579 $\pm$ 0.004 & 11.368 $\pm$ 0.004 & 11.007 $\pm$ 0.004 & -1.77 & 0.014 \\
		SW And & RRab & 0.442 & 1.995 $\pm$ 0.031 & 10.097 $\pm$ 0.006 & 9.692 $\pm$ 0.006 & 9.169 $\pm$ 0.008 & -0.24 & 0.038 \\
		SW Dra & RRab & 0.570 & 1.056 $\pm$ 0.020 & 10.815 $\pm$ 0.006 & 10.470 $\pm$ 0.006 & 9.977 $\pm$ 0.006 & -1.12 & 0.014 \\
		SX UMa & RRc & 0.307 & 0.882 $\pm$ 0.015 & 11.04 $\pm$ 0.01 & 10.847 $\pm$ 0.01 & 10.532 $\pm$ 0.01 & -1.81 & 0.010 \\
		T Sex & RRc & 0.325 & 1.324 $\pm$ 0.025 & 10.294 $\pm$ 0.008 & 10.032 $\pm$ 0.008 & 9.673 $\pm$ 0.008 & -1.34 & 0.044 \\
		TT Lyn & RRab & 0.597 & 1.464 $\pm$ 0.017 & 10.217 $\pm$ 0.011 & 9.853 $\pm$ 0.011 & 9.318 $\pm$ 0.011 & -1.56 & 0.017 \\
		TU UMa & RRab & 0.558 & 1.593 $\pm$ 0.028 & 10.156 $\pm$ 0.006 & 9.816 $\pm$ 0.006 & 9.3189 $\pm$ 0.006 & -1.51 & 0.022 \\
		TV Boo & RRc & 0.313 & 0.756 $\pm$ 0.017 & 11.179 $\pm$ 0.009 & 10.985 $\pm$ 0.009 & 10.652 $\pm$ 0.009 & -2.44 & 0.010 \\
		TW Her & RRab & 0.400 & 0.901 $\pm$ 0.016 & 11.554 $\pm$ 0.005 & 11.249 $\pm$ 0.005 & 10.817 $\pm$ 0.005 & -0.69 & 0.042 \\
		UU Vir & RRab & 0.476 & 1.266 $\pm$ 0.052 & 10.868 $\pm$ 0.006 & 10.561 $\pm$ 0.006 & 10.118 $\pm$ 0.006 & -0.87 & 0.018 \\
		UV Oct & RRab & 0.543 & 1.834 $\pm$ 0.013 & 9.844 $\pm$ 0.006 & 9.471 $\pm$ 0.006 & 8.94 $\pm$ 0.006 & -1.74 & 0.091 \\
		UY Boo & RRab & 0.651 & 0.809 $\pm$ 0.025 & 11.28 $\pm$ 0.009 & 10.927 $\pm$ 0.009 & 10.427 $\pm$ 0.009 & -2.56 & 0.033 \\
		UY Cam & RRc & 0.267 & 0.757 $\pm$ 0.020 & 11.685 $\pm$ 0.009 & 11.507 $\pm$ 0.009 & 11.207 $\pm$ 0.027 & -1.33 & 0.022 \\
		UY Cyg & RRab & 0.561 & 0.953 $\pm$ 0.023 & 11.515 $\pm$ 0.017 & 11.095 $\pm$ 0.017 & 10.495 $\pm$ 0.017 & -0.80 & 0.129 \\
		V Ind & RRab & 0.480 & 1.490 $\pm$ 0.021 & 10.282 $\pm$ 0.006 & 9.9720 $\pm$ 0.006 & 9.509 $\pm$ 0.006 & -1.50 & 0.043 \\
		V0440 Sgr & RRab & 0.477 & 1.390 $\pm$ 0.023 & 10.703 $\pm$ 0.008 & 10.312 $\pm$ 0.008 & 9.805 $\pm$ 0.008 & -1.40 & 0.085 \\
		V0675 Sgr & RRab & 0.642 & 1.184 $\pm$ 0.021 & 10.706 $\pm$ 0.006 & 10.298 $\pm$ 0.006 & 9.72 $\pm$ 0.006 & -2.28 & 0.130 \\
		VX Her & RRab & 0.455 & 1.040 $\pm$ 0.030 & 10.978 $\pm$ 0.009 & 10.689 $\pm$ 0.009 & 10.244 $\pm$ 0.009 & -1.58 & 0.044 \\
		W Crt & RRab & 0.412 & 0.841 $\pm$ 0.029 & 11.844 $\pm$ 0.006 & 11.530 $\pm$ 0.006 & 11.099 $\pm$ 0.006 & -0.54 & 0.040 \\
		WY Ant & RRab & 0.574 & 0.978 $\pm$ 0.023 & 11.217 $\pm$ 0.004 & 10.850 $\pm$ 0.004 & 10.324 $\pm$ 0.004 & -1.48 & 0.059 \\
		X Ari & RRab & 0.651 & 1.868 $\pm$ 0.021 & 10.061 $\pm$ 0.006 & 9.5620 $\pm$ 0.006 & 8.868 $\pm$ 0.006 & -2.43 & 0.180 \\
		XX And & RRab & 0.723 & 0.889 $\pm$ 0.022 & 11.017 $\pm$ 0.009 & 10.675 $\pm$ 0.009 & 10.145 $\pm$ 0.009 & -1.94 & 0.039 \\
		XZ Cyg & RRab & 0.467 & 1.584 $\pm$ 0.016 & 9.9029 $\pm$ 0.008 & 9.645 $\pm$ 0.008 & 9.237 $\pm$ 0.008 & -1.44 & 0.096 \\
		YZ Cap & RRc & 0.273 & 0.883 $\pm$ 0.034 & 11.56 $\pm$ 0.006 & 11.3 $\pm$ 0.006 & 10.902 $\pm$ 0.006 & -1.06 & 0.063 \\
		\enddata
	\end{deluxetable*}

	\clearpage
	\mbox{}
	\clearpage
	\section{Application to Mock Galaxies}
	\label{sec:mockdata}
	
	Here, we verify our model performance using mock data of three galaxies. We set $N = 2$, $N = 31$, and $N = 263$ RRab stars for Galaxy 0, Galaxy 1, and Galaxy 2, respectively.   We assigned random, realistic distance moduli for each galaxy: $\mu_0 = 24.92$, $\mu_1 = 21.85$, and $\mu_2 = 19.97$. We also assigned MDFs to each galaxy using the MDFs listed in  Table \ref{tab:mock_priors}. For each star, we generated a unique log period value using the prior in Table \ref{tab:mock_priors}. Next, in order to generate apparent Wesenheit magnitudes for each star, we assumed one of the following two PWZ models:
	
	\begin{equation}
	\begin{split}
	W_{i, j} = & -1.00 - 1.8 \log{P_{i, j}} + 0.25 {\rm [Fe/H]}_{i, j} \\ & + \mu_i + \sigma_{\rm noise} + \sigma_{\rm intr}
	\end{split}
	\end{equation}
	
	\begin{equation}
	\begin{split}
	W_{i, j} = & -1.00 - 1.8 \log{P_{i, j}} + 0.15 {\rm [Fe/H]}_{i, j} \\ & + \mu_i + \sigma_{\rm noise} + \sigma_{\rm intr},
	\end{split}
	\end{equation}
	where we modeled the observational error $\sigma_{\rm noise}$ and the intrinsic scatter $\sigma_{\rm intr}$ using the respective distributions given in Table \ref{tab:mock_priors}. Finally, we created a batch of 50 calibration stars, intended to replicate the stabilizing effects of the {\it Gaia} parallax values for the Milky Way field RR Lyrae. The apparent Wesenheit magnitudes for these stars were calculated using the same PWZ model as the other stars, but with a larger noise term:
	
	\begin{equation}
	\sigma_{\rm noise}^{'} \sim \mathcal{N}(0, 0.102),
	\end{equation}
	The increased noise accounts for an additional error of 0.1 mag, associated with the uncertainty in the distance modulus of each calibration star, added in quadrature to the photometric error of 0.02 mag.
	
\begin{figure*}[h!]
		\centering
		\includegraphics[width = \textwidth]{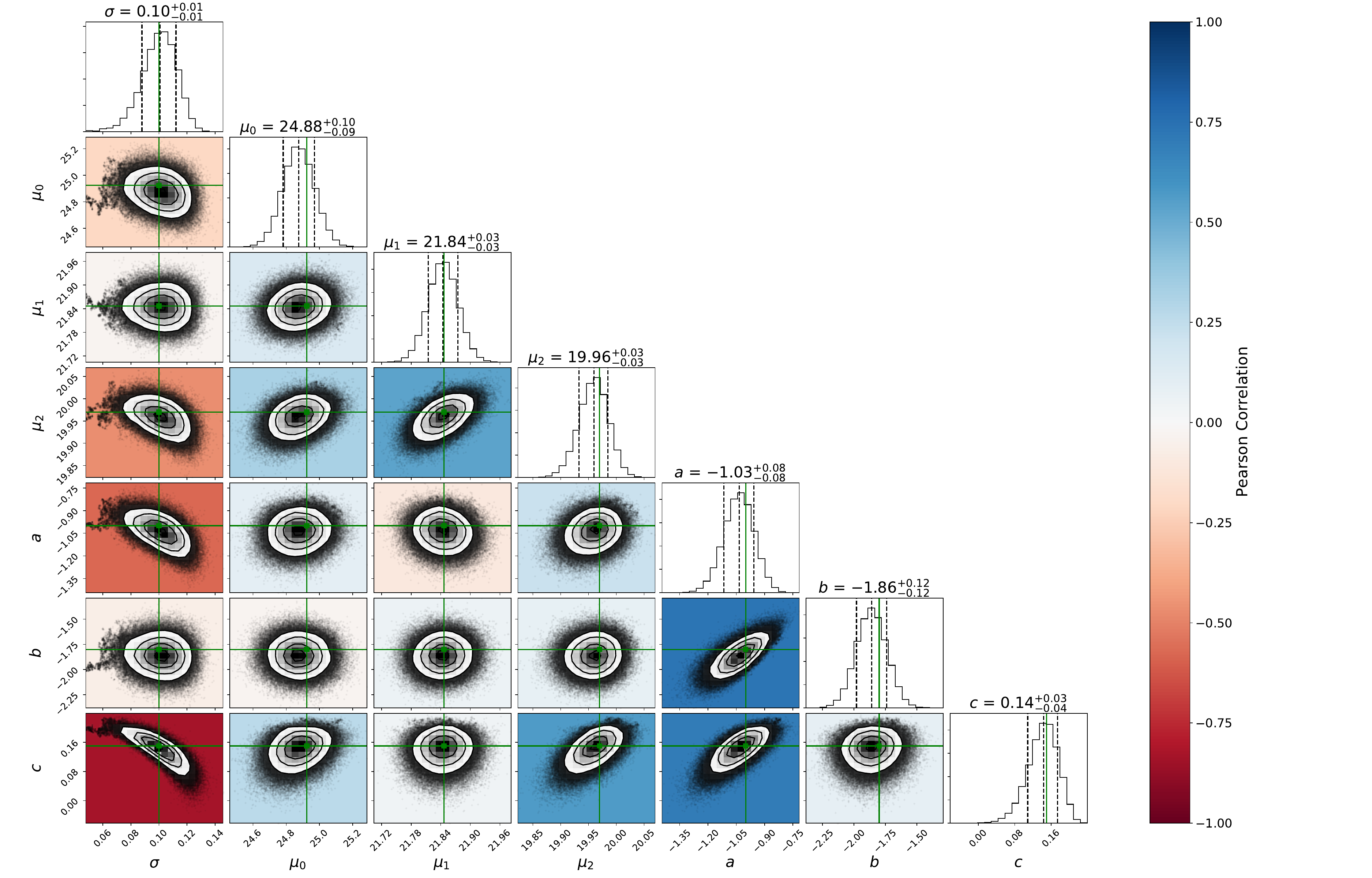}
		\caption{Example of a corner plot from one of our mock data trials, which assumed a metallicity slope of 0.15 and used default MDF priors. The vertical green lines indicate the truth values. The diagonal entries display the marginal distribution of each derived parameter, with dashed lines indicating the median value and 1-sigma cutoffs. Each of the other panels displays a joint distribution, with contours identifying regions of high sampling density in parameter space. Dark red panels indicate strong negative correlation values, while dark blue panels indicate strong positive correlation values.}
		\label{fig:mock_corner}
	\end{figure*}

	Figure \ref{fig:mock_corner} shows the recovered parameters from running the mock galaxies through our model.  In all cases, the input parameters are recovered within 1-$\sigma$.
	
\begin{figure*}[h!]
		\centering
		\includegraphics[width = \textwidth]{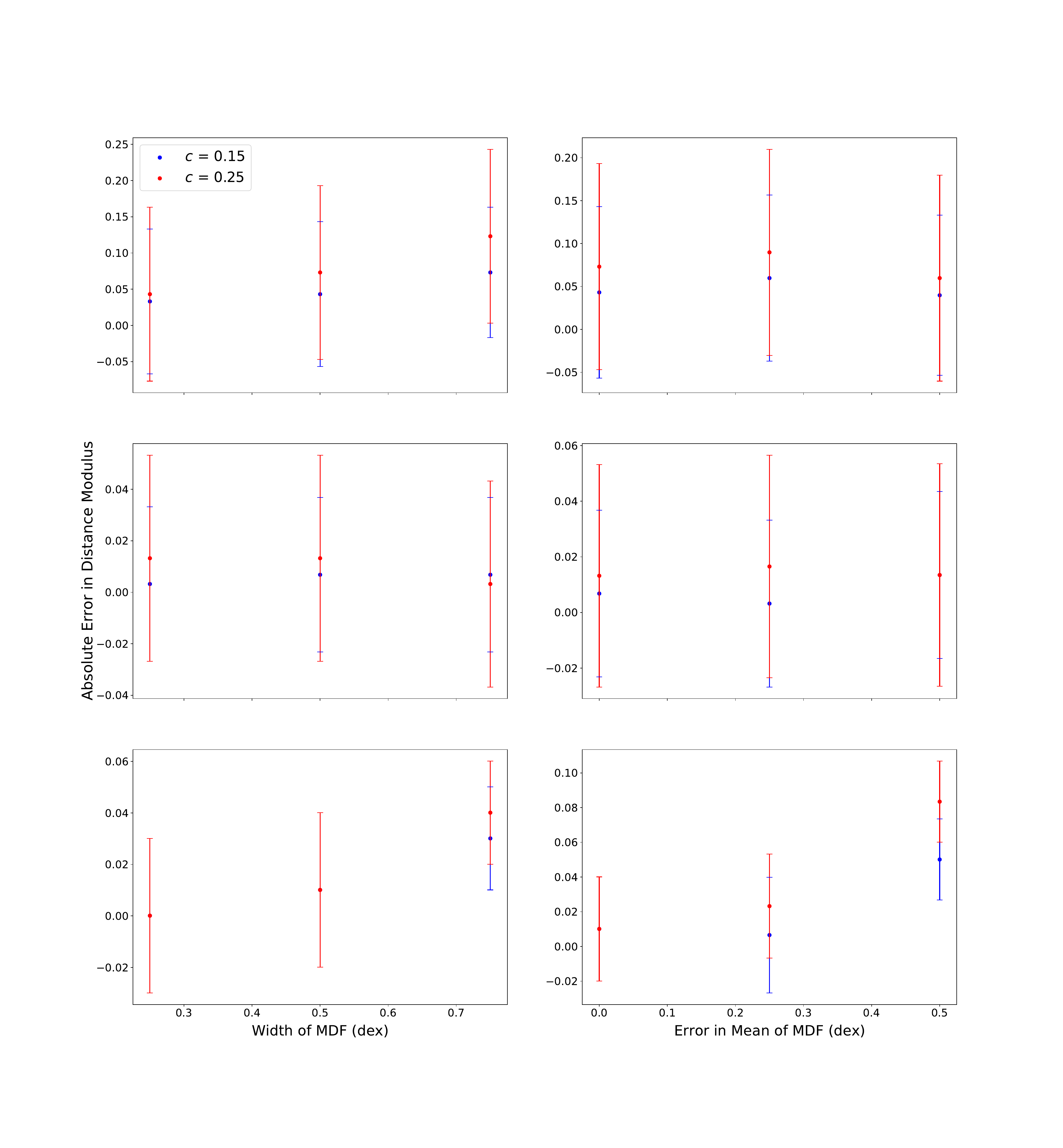}
		\caption{Sensitivity of recovered distance modulus of mock galaxies to perturbations in the standard deviation and mean of the MDF prior. In the left column, the scatter plots display average errors in recovered distance moduli for an underestimated MDF width of 0.25 dex, a true MDF width of 0.5 dex, and an overestimated MDF width of 0.75 dex. In the right column, the scatter plots display average errors in recovered distance moduli for perturbations of 0 dex, 0.25 dex, and 0.5 dex to the mean metallicity. In both panels, results are reported for metallicity slopes of 0.15 and 0.25, and the plotted error bars indicate the typical uncertainty in the derived distance modulus for each trial.}
		\label{fig:mock_trials}
	\end{figure*}

	As an additional test, we check the sensitivity of parameter recovery to  perturbations in the MDF of each galaxy.  First, we changed the MDF prior to 0.25 dex and 0.75 dex (as opposed to the true value of 0.5 dex). As shown in Figure \ref{fig:mock_trials}, these changes affect the distance recovery by $\sim1$\%.  We found that only large changes (0.75 dex; Galaxy 2) in the MDF width led to a $\sim3$\% bias in recovered distance.  These results do not depend strongly on the adopted PWZ metallicity slope.
	
	We next checked the effect of having an inaccurate assumed mean metallicity by running the model on galaxies with 0.25 dex and 0.5 dex biases imposed on the mock data.  As shown in the right panel of Figure \ref{fig:mock_trials}, we found that the 0.25~dex offset had modest effects on the recovered distances, whereas the 0.5~dex offset introduced a bias of $0.05$-$0.09$ in the recovered distance, depending on the number of stars in the galaxy.

	\begin{deluxetable}{ccc}
		\tablecaption{Prior distributions used in the MCMC model for mock data.}
		\label{tab:mock_priors}
		\tablehead{\colhead{Parameter} &
			\colhead{Prior} & \colhead{Description}}
		\startdata
		MDF$_0$ & $\mathcal{N}$(-2.35, 0.5) & MDF of Galaxy Zero \\
		MDF$_1$ & $\mathcal{N}$(-1.56, 0.5) & MDF of Galaxy One \\
		MDF$_2$ & $\mathcal{N}$(-2.00, 0.5) & MDF of Galaxy Two \\
		$\log_{10}{P}$ & $\mathcal{N}$(-0.35, -0.55) & RR Lyrae Period \\
		$\sigma_{\rm noise}$ & $\mathcal{N}$(0, 0.02) & Observational Error \\
		$\sigma_{\rm intr}$ & $\mathcal{N}$(0, 0.1) & Intrinsic Scatter \\
		\enddata
	\end{deluxetable}

	\section{Corner Plots for Model Fits to All Galaxies}

\begin{figure*}[h!]
		\centering
		\includegraphics[width = \textwidth]{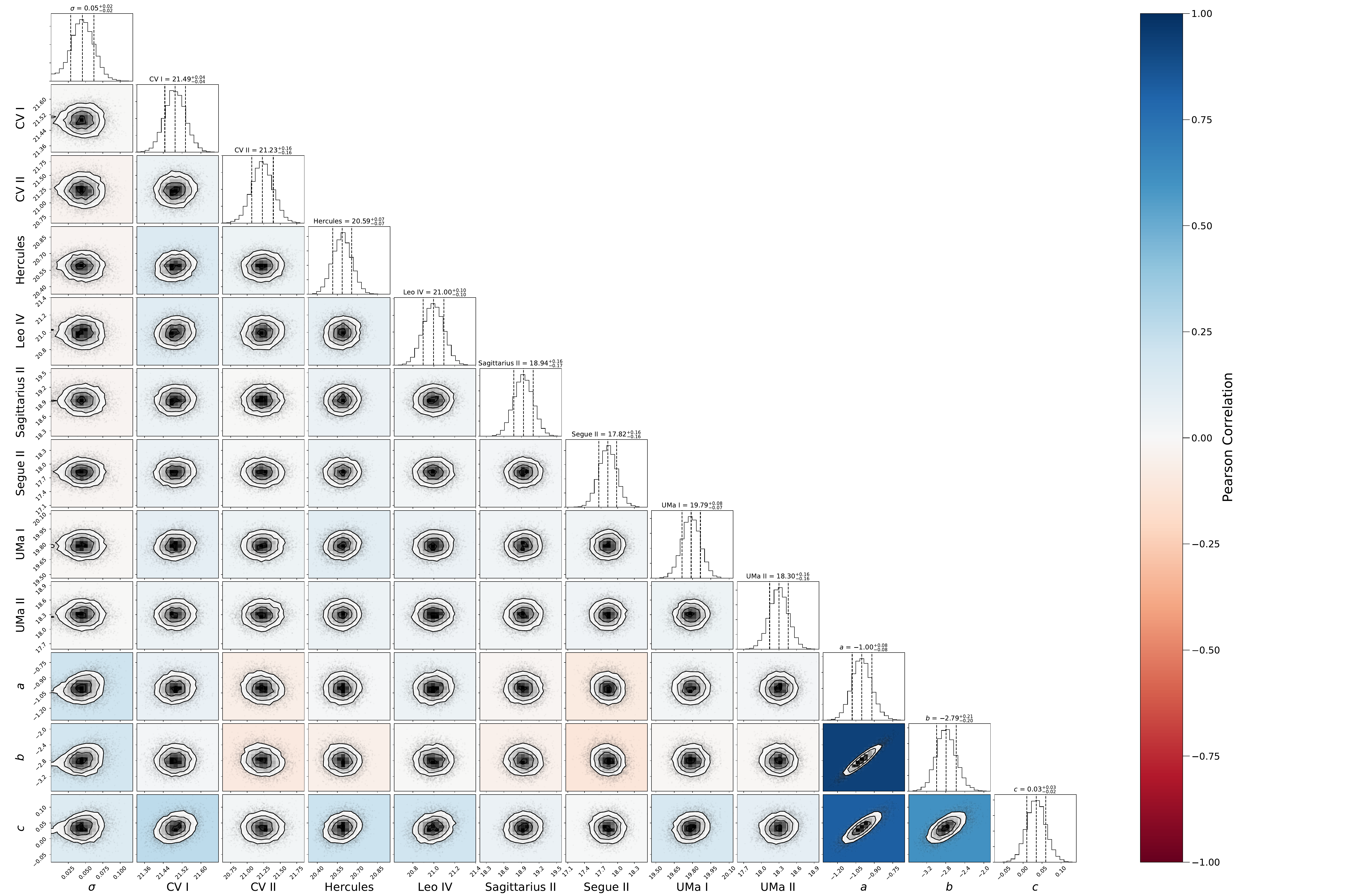}
		\caption{Full corner plot from the final run of our MCMC model for B-V Wesenheit Magnitude galaxies. The diagonal entries display the marginal distribution of each derived parameter, with dashed lines indicating the median value and 1-sigma cutoffs. Each of the other panels displays a joint distribution, with contours identifying regions of high sampling density in parameter space.}
		\label{fig:cornerBV}
	\end{figure*}

	Figures \ref{fig:cornerVI} and \ref{fig:cornerBV} show show 
	corner plots from our MCMC runs for the V-I and B-V full sample of galaxies.  A detailed description of the plot layout and discussion of general trends is presented in \S\S \ref{sec:method} and \ref{sec:results}.

	\begin{figure*}[h!]
		\centering
		\includegraphics[width = \textwidth]{Final_Corner_Plot_V-I.pdf}
		\caption{Full corner plot from the final run of our MCMC model for V-I Wesenheit Magnitude galaxies. The diagonal entries display the marginal distribution of each derived parameter, with dashed lines indicating the median value and 1-sigma cutoffs. Each of the other panels displays a joint distribution, with contours identifying regions of high sampling density in parameter space. }
		\label{fig:cornerVI}
	\end{figure*}

	\clearpage
	

	\bibliographystyle{aasjournal}
	
	
	
\end{document}